\DocumentMetadata{pdfstandard=A-3u,pdfversion=1.7,lang=en}
\documentclass[
    subscriptcorrection,
    upint,
    varvw,
    barcolor=Goldenrod3,
    mathalfa=cal=euler,
    balance,
    hyphenate,
    french
]{asmejour} %
\usepackage[normalem]{ulem}

\hypersetup{%
    pdfauthor={Jordan D. Roberts, Vibodha Bandara, Kenichi Nakano, Dustin Keller}, 
    pdftitle={An ASME-Compliant Helium-4 Evaporation Refrigerator for the SpinQuest Experiment}, 
    pdfkeywords={Spin physics, polarized target, helium refrigeration, cryogenics, process piping, ASME B31.3, ASME BPVC, FESHM compliance}, 
    pdfsubject={Design of a compliant helium-4 evaporation refrigerator for the SpinQuest experiment at Fermilab. }, 
}

\JourName{Pressure Vessel Technology}

\begin{document}

    \SetAuthorBlock{Jordan D. Roberts\CorrespondingAuthor}{Department of Mechanical and Aerospace Engineering,\\ University of Virginia,\\ 122 Engineer's Way,\\ Charlottesville, VA, USA \\ email: ffk9mr@Virginia.edu}


    \SetAuthorBlock{Vibodha Yasas Sri Bandara}{Department of Instrumentation and Automation Technology, \\ University of Colombo,\\ Colombo, Sri Lanka,\\ email: vibodha@iat.cmb.ac.lk }

    \SetAuthorBlock{Kenichi Nakano}{Department of Physics, \\ University of Virginia,\\ 382 McCormick Rd\\ Charlottesville, VA, USA \\ email: bhy7tf@virginia.edu }

    \SetAuthorBlock{Dustin Keller}{Department of Physics, \\ University of Virginia,\\ 382 McCormick Rd\\ Charlottesville, VA, USA \\ email: dustin@Virginia.edu }

    \title{An ASME-Compliant Helium-4 Evaporation Refrigerator for the SpinQuest
    Experiment}

    \keywords{Spin physics, polarized target, helium refrigeration, cryogenics,
    process piping, ASME B31.3, ASME BPVC, FESHM compliance.}


    \begin{abstract}
        This paper presents the design, safety basis, and commissioning results for
        a 1~K liquid helium-4 (\textsuperscript{4}He) evaporation refrigerator developed
        for the Fermi National Accelerator Laboratory (FNAL) SpinQuest experiment
        (E1039). The system is the first high-power helium evaporation refrigerator
        operated in a fixed-target scattering experiment at FNAL. It was engineered
        to satisfy FNAL Environment, Safety, and Health Manual (FESHM) requirements
        for pressure vessels, piping, cryogenic systems, and vacuum vessels. The
        pressure-boundary design and overpressure protection were mapped to ASME
        B31.3 and the ASME Boiler and Pressure Vessel Code (BPVC), with documented
        compliance with FESHM Chapters~5031, 5031.1, 5032, 5033, and 5064. We
        describe a transferable compliance methodology for high-power
        \textsuperscript{4}He evaporation refrigerators, including pressure analysis,
        relief sizing, fabrication traceability, and operational review. The design also mitigates the high-radiation target-cave environment by locating the active (transistor-based) electronics approximately 20 m from the cryostat, while maintaining stable pressure, flow, and temperature control through connections to the passive mechanical components in the cave. Commissioning data from July~2024
        demonstrate stable operation near 1~K under beam and dynamic nuclear
        polarization microwave loading, successful thermal-equilibrium control, and
        safe performance of the relief network during transient events.
    \end{abstract}

    \date{Version \versionno, \today}

    \maketitle 

\section{Introduction}
    Solid polarized targets entered scattering experiments in the early 1960s. The enabling technique was dynamic nuclear polarization (DNP), in which nuclear polarization is produced in a solid by coupling the nuclear spin system to paramagnetic centers at low temperature and high magnetic field. Abragam's work established the DNP principle \cite{abragam_cea,abragam}, and in 1962 Abragam, Borghini, and collaborators at Saclay reported the scattering of 20~MeV polarized protons from a polarized proton target based on a lanthanum magnesium nitrate (LMN) crystal, measuring the spin-correlation parameter $C_{nn}$ \cite{ABRAGAM1962310}. Chamberlain, Jeffries, Schultz, Shapiro, and Van Rossum soon extended the method to higher-energy particle scattering at Berkeley with a polarized proton target for 246~MeV $\pi^+$--proton scattering \cite{Chamberlain1963}. Borghini's later spin-temperature description and target-material developments provided a practical framework for accelerator-based DNP targets \cite{MB,Borghini2}. These early systems established the architecture still recognizable in modern polarized targets: a strong holding field, a low-temperature refrigerator, microwave-driven DNP, and NMR-based polarization monitoring.

    Through the 1970s and 1980s, polarized targets became central to spin-dependent lepton--nucleon and hadron--nucleon scattering. A major milestone was the European Muon Collaboration (EMC) polarized muon--proton deep-inelastic scattering measurement at CERN, which determined the proton spin-dependent structure function $g_1^p$ and helped initiate the modern proton-spin program \cite{EMC1988,Ashman1989}. This work preceded and motivated later high-luminosity polarized-target experiments at BNL \cite{3241,bnl}, SLAC \cite{E143,E155}, and Jefferson Lab \cite{clas,g2p,APOLO,E03,e06,E93,SANE,EG1,RAS}. Many of these later systems used Crabb-style solid polarized targets \cite{E143} containing materials such as \textsuperscript{6}LiD, NH\textsubscript{3}, or ND\textsubscript{3}.

    The cryogenic requirements of these targets evolved with the experimental program. Early polarized-target experiments emphasized low temperature and high polarization in relatively small target volumes. As beam intensity, target size, and DNP microwave power increased, the limiting requirement shifted toward removing beam- and microwave-induced heat while preserving polarization near 1~K. For high-intensity fixed-target operation, \textsuperscript{4}He evaporation refrigerators became especially important because they provide watt-scale cooling near 1~K, roughly two to three orders of magnitude greater than typical \textsuperscript{3}He evaporation or dilution systems used in earlier polarized-target applications \cite{Crabb}. The refrigerator therefore became not only a low-temperature device but also the central thermal-management element coupling beam delivery, DNP operation, target polarization, and experiment uptime.

    During the 6~GeV era at Jefferson Lab's Continuous Electron Beam Accelerator Facility (CEBAF), two polarized solid targets were successfully operated: the University of Virginia (UVa) target and the Hall B target.
    The UVa target was the first polarized solid target used at Jefferson Lab. It first supported the $G_n^e$ experiment in 1998 with a 2.725~GeV electron beam at 100~nA \cite{GnE}\cite{inproceedings}. This target had previously been employed at SLAC for several high-energy experiments (up to 29.7~GeV) with beam intensities of $1$--$4 \times 10^9$ electrons per pulse \cite{SLAC2,E143,E155}. It was the standard target system for several experiments at Jefferson lab and was later fitted with an upgraded evaporation refrigerator with other renovations for the $g_2^p$ and $G_E^p$ experiments in Hall A and to comply with Jefferson Lab's ASME pressure-vessel guidelines. The upgraded system achieved a cooling power of approximately 3~W at 1.4~K using a 12,000~m$^3$/h roots blower pump stack \cite{g2p}.
    The Hall B target utilized a Roubeau-style horizontal evaporation refrigerator that provided 0.8~W of cooling power at 1.1~K, backed by a 3,300~m$^3$/hr roots pump stack \cite{EG1}. Due to the tight geometric constraints of the CLAS detector, which limited insulation, the refrigerator's performance was restricted; in an ideal configuration the design was capable of reaching approximately 1.5~W at 1.1~K \cite{ROUBEAU}.

    With the completion of the 12~GeV upgrade of CEBAF, Jefferson Lab now benefits from additional polarized target options and improved capabilities, particularly for the CLAS12 detector in Hall B and other ongoing experiments \cite{PandeyClas12}. 

    The recent polarized target history demonstrates that refrigerator designs are strongly coupled to the experiment's configuration needs. As beam intensities increase and geometric constraints are relaxed, higher cooling power becomes achievable. However, a comprehensive end-to-end design and compliance guide for such systems remains to be fully developed.
    
    Large DNP target systems also introduce pressure-system challenges. The helium flow paths that enable 1~K operation must withstand off-normal events such as loss of insulating vacuum, blocked flow, rapid warm-up, and magnet-quench transients. Higher mass flow increases the demands on transfer lines, pumping paths, helium vessels, and relief devices, while transient heat loads from DNP microwaves and pulsed or continuous beams can increase the target heat load. Consequently, the refrigerator, pumping stack, relief network, and instrumentation cannot be treated as independent subsystems; together they form a coupled cryogenic and pressure-safety system.

    During the same period, pressure-vessel and cryogenic safety standards for these systems have become increasingly formal. U.S. Department of Energy (DOE) laboratories now enforce institutional and national pressure-system requirements more rigorously than in earlier polarized-target programs. Fermi National Accelerator Laboratory (FNAL) is representative: the Fermilab Environment, Safety, and Health Manual (FESHM) \cite{fes} invokes the ASME Boiler and Pressure Vessel Code (BPVC) Section~VIII \cite{asme81} and ASME B31.3 Process Piping Code \cite{b31} for credited pressure systems. These codes are essential for pressure-boundary safety, but their nominal assumptions are directed toward industrial pressure equipment operating at ambient or elevated temperature. They do not map directly onto thin-walled, low-pressure cryogenic vessels that spend their service life near 4~K. Applying them to polarized-target refrigerators therefore requires conservative design choices and explicit engineering justification for thermal contraction, low-temperature material behavior, transient heat loading, weld qualification, inspection, relief sizing, and system-level approval \cite{Gentzlinger}.

    SpinQuest (E1039) at FNAL brings these requirements together. The experiment requires a high-cooling-power \textsuperscript{4}He evaporation refrigerator that operates inside the bore of a 5~T superconducting magnet, maintains a polarized target near 1~K, and absorbs rapidly varying heat loads from DNP microwaves and a high-intensity 120~GeV proton beam delivering more than 10\textsuperscript{13} protons per 4~s spill on a 60~s cycle. The hadron-beam target cave also creates a radiation environment that precludes conventional local electronic instrumentation near the cryostat. The refrigerator must therefore combine watt-class cryogenic performance, remote radiation-tolerant instrumentation, and a documented pressure-code safety basis satisfying FESHM, BPVC Section~VIII, and B31.3.

    Prior work at CERN and Jefferson Lab has shown that ASME-based approaches can be applied to cryogenic refrigeration and accelerator systems \cite{g2p,Gentzlinger,clas12,inproceedings,HallC,APOLO}. However, the published literature does not document an end-to-end methodology for achieving pressure-code compliance in a high-power \textsuperscript{4}He evaporation refrigerator operating in a high-radiation hadron-beam environment. Existing references address pressure-boundary design, overpressure protection, or remote instrumentation as separate topics, but they do not integrate all three into a single safety basis that simultaneously satisfies BPVC Section~VIII, B31.3, and laboratory-specific FESHM approval processes. This gap forces each new experiment to reconstruct the integration from primary sources, lengthens design-review cycles, and leaves reviewers without a documented benchmark for comparable systems.

    To address this gap, the UVa Spin Physics Group designed, built, and commissioned a \textsuperscript{4}He evaporation refrigerator for SpinQuest. The system was developed to meet the thermal demands of DNP operation under a high-intensity proton beam while satisfying FNAL pressure-code and cryogenic safety requirements. The design uses substantial margins relative to ASME code allowables, redundant overpressure protection, and radiation-tolerant remote instrumentation located outside the target cave. At FNAL, this system is the first \textsuperscript{4}He evaporation refrigerator operated in a fixed-target scattering experiment.

    This paper presents the design methodology, pressure-code safety basis, and commissioning results for the UVa \textsuperscript{4}He evaporation refrigerator built for SpinQuest. We validate thermal performance through direct heat-load and cooling-capacity measurements, demonstrate operational stability during magnet-quench testing, and report thermal-equilibrium data obtained during steady-state operation. Together, these results establish an end-to-end, FESHM-compliant design and qualification methodology for watt-class \textsuperscript{4}He evaporation refrigerators under high-intensity hadron-beam conditions and provide a documented benchmark for future DNP target systems at Fermilab and comparable laboratories.

    The remainder of this paper is organized as follows. Section~\ref{sec:barriers} identifies the key technical barriers. Section~\ref{code} presents the code-compliance framework. Section~\ref{system_overview} describes the system design. Section~\ref{Control} details the measurement and control architecture. Section~\ref{commissioning} presents the commissioning results, and Section~\ref{conclusion} gives the conclusions and implications for future high-intensity polarized-target systems.

\section{Key Technical Barriers}
\label{sec:barriers}
    Although many individual components follow established engineering practice, integrating them in a high-radiation, high-heat-load, code-constrained environment introduces coupled requirements that standard references do not resolve. This work addresses three primary barriers: applying ASME pressure-vessel and piping codes to a low-pressure cryogenic refrigerator, achieving reliable remote instrumentation in a high-radiation environment, and integrating high cooling power with transient heat-load management and overpressure protection.

    \subsection{Application of ASME Pressure-Vessel and Piping Codes to Low-Pressure Cryogenic Systems}
    ASME BPVC Section~VIII \cite{asme81} was written principally for industrial pressure vessels operating above atmospheric pressure at ambient or elevated temperature. Its provisions do not map cleanly onto a cryogenic system that operates near atmospheric pressure and below 4~K. The code's allowable-stress tables, overpressure-protection provisions, and classification rules assume a service regime that the SpinQuest refrigerator does not occupy, and FESHM \cite{fes} inherits that ambiguity when it invokes Section~VIII for experiment-specific pressure equipment. We resolve the ambiguity with a compliance methodology that combines Section~VIII design-by-analysis with FESHM vessel requirements: conservative allowable stresses derived from low-temperature material data, relief devices sized against the bounding loss-of-vacuum and magnet-quench scenarios, and full pressure-boundary traceability from design through fabrication and test. This methodology establishes a defensible safety basis for a pressure system that would otherwise fall in a gray area of code applicability. Section~\ref{code} outlines the code and laboratory requirements that frame the design, and Section~\ref{system_overview} presents the resulting system architecture.

    \subsection{Reliable Cryogenic Operation in a High-Radiation Environment with Remote Instrumentation}
    The SpinQuest beam environment delivers total ionizing doses sufficient to cause rapid failure of conventional semiconductor instrumentation placed near the cryostat \cite{semirad}. Standard cryogenic control electronics therefore cannot be deployed locally. All active instrumentation must be placed approximately 20~m from the refrigerator, while pressure, flow, and temperature control must remain accurate and stable. We address this constraint through three complementary measures: passive sensing lines referenced to radiation-shielded transducers, radiation-tolerant components at the limited sensor locations that must remain in the beam area, and remotely actuated control valves with mechanical position feedback. Together, these measures support stable 1~K operation and closed-loop control from outside the radiation field. Section~\ref{Control} presents the instrumentation architecture and its measured performance.

    \subsection{Integration of High Cooling Power with Transient Heat Loads and Overpressure Protection}
    Thermal transients, pressure excursions, and credible failure modes are tightly coupled in the SpinQuest refrigerator. During normal operation, rapidly varying heat loads from the environment, DNP microwaves, and 120~GeV proton-beam spills drive bursts of helium boil-off that propagate into the vapor space as pressure excursions. The refrigerator must absorb those transients while holding the target near 1~K. Under credible off-normal conditions, a magnet quench or vacuum-jacket failure can generate helium vapor at rates far above the capacity of the primary pumping path; the relief network must therefore handle the excess flow. We address this coupled thermal-fluid and safety problem through coordinated design of the pumping system, flow-impedance distribution, and relief network, with relief devices sized to the overpressure-protection provisions of BPVC Section~VIII and the venting requirements of FESHM \cite{fes,asme81}. Section~\ref{code} details these provisions. A comprehensive quench study will be reported separately \cite{quench}. Section~\ref{commissioning} presents heat-load accounting for the environmental, microwave, and beam contributions, together with the cooling power derived from the main helium flow.

    Together, these three barriers define the engineering scope of this work. Resolving them required an integrated design methodology that jointly addresses cryogenic performance, radiation-hardened instrumentation, and pressure-system code compliance.

\section{Codes, Standards, and Laboratory Requirements}
\label{code}

    Although the SpinQuest refrigerator operates near atmospheric pressure, it was evaluated systematically against ASME BPVC~\cite{asme81}, ASME B31.3~\cite{b31}, and FESHM~\cite{fes}. This section summarizes the compliance basis, including maximum allowable working pressure (MAWP) calculations, relief analysis, and operational safety margins.

    The Fermilab Environment, Safety, and Health Manual (FESHM) \cite{fes} requires pressure vessels either to comply with ASME BPVC~\cite{asme81} or to be classified as ``Experiment Vessels.'' Experimental vessels are subject to additional requirements, including a 0.8 reduction factor on BPVC allowable stresses, Section~IX-qualified weld and braze procedures, and an engineering note verifying overpressure protection \cite{fes}. The applicable FESHM chapter determines the required review path, and best engineering practice is enforced throughout the design and approval process.

    \subsection{FESHM Safety Procedures}
    The relevant FESHM chapters define procedures required for safe operation of the \textsuperscript{4}He evaporation refrigerator. These include Work Planning and Control (WPC), Operational Readiness Clearance (ORC), engineering notes, and cryogenic review. The engineering note process is especially important because it provides the traceable technical basis for pressure-system approval.

    \subsubsection{WPC and ORC}
    The WPC process defines the task scope, hazards, authorization, performance requirements, and evaluation criteria for the work \cite{fes}. Formal plans include written hazard analysis, permits, inspection or walk-through requirements, and operating procedures. Operation of the equipment also requires an ORC review, a graded committee review that verifies that ES\&H issues are properly managed \cite{fes}. Each component required a separate WPC and ORC review before installation in the experimental hall.

    \subsubsection{Engineering Note}
    An engineering note is a written analysis, prepared by a qualified person, demonstrating that a vessel satisfies the requirements of the applicable FESHM chapter \cite{fes}. For example, FESHM~5031 \cite{fes} specifies three documentation categories that are standard across most engineering notes:
    \begin{itemize}
        \item \textbf{Identification and description:} formal identification of the part; description of the vessel, purpose, and installation location; and the method used to establish the MAWP. Supporting material can include drawings, bills of material, technical reports, and piping and instrumentation diagrams.
        \item \textbf{System venting verification:} relief analysis and heat-transfer studies establishing relief size, capacity, set pressure, test procedure, and stored energy, with compliance evaluated under applicable API guidance~\cite{API}.
        \item \textbf{Operating procedures:} risk assessments, operating procedures, and required worker training for all credible hazards.
    \end{itemize}

    \subsubsection{Cryogenic Review}
    FESHM~5032 defines the cryogenic review process and occupational training requirements for systems operating below $-150^\circ$C, including refrigerated magnets, accelerating structures, and cryogenic targets. The classification also covers warm systems connected to cryogenic systems through fluid lines and used to capture, store, or process gas, as well as cryogenic purge-gas supplies for stored liquid inventories above 200~L. FNAL convenes a Cryogenic Safety Panel (CSP) to review cryogenic safety and determine the protocols required for compliance. This CSP review is a standard FESHM requirement for FNAL-integrated cryogenic systems.

\subsection{Fabrication and ODH Controls}

    All welded components were fabricated by ASME-certified welders. Leak integrity was verified by helium mass-spectrometer testing, with a minimum measured leak rate of $9.8 \times 10^{-9}$~std~cm$^{3}$/s. Because cryogenic helium systems present an oxygen-deficiency hazard (ODH), the refrigerator area is equipped with oxygen monitors, alarms, and ventilation. ODH classification and mitigation followed FESHM~5064 \cite{fes}, and SpinQuest personnel are trained in Fermilab ODH procedures.

\subsection{Compliance Traceability Matrix}
\label{matrix}
    The refrigerator design was evaluated against applicable codes and Fermilab safety standards. Internal engineering notes document each calculation to Fermilab standards \cite{uva-note,fnal-parallel-plate,sep-relief,lhe-relief,quench,window}. Piping systems were designed to ASME B31.3~\cite{b31} for Category~D helium service~\cite{asme2}. Pressure-vessel calculations followed ASME BPVC Section~VIII, Division~1 \cite{asme81} for MAWP, wall thickness, and material selection. Overpressure-protection principles from BPVC Sections~VIII and XIII were followed, with parallel-plate vent capacities informed by Fermilab test data and CGA\,S-1.3 \cite{cga}. Relief sizing followed API~520 \cite{API}. Table~\ref{tab:compliance} summarizes the compliance framework.

    \begin{table}[h!]
        \centering
        \caption{Compliance traceability matrix.}
        \label{tab:compliance}
        \begin{tabular}{p{0.2\linewidth} p{0.7\linewidth}}
            \toprule \textbf{Standard} & \textbf{Application in Refrigerator Design} \\
            \midrule ASME B31.3        & Tubing and piping design for Category~D helium service; material allowables (Table~A-1); minimum wall thickness (Para.~304.1.2). \\
            ASME BPVC VIII-1           & MAWP calculations for vessels (UG-27, UG-28, UG-34); welding qualifications (Section~IX); overpressure-protection principles (UG-125--UG-137). \\
            API\,520                  & Relief sizing. \\
            CGA\,S-1.3                 & Relief sizing. \\
            FESHM 5031                 & Experiment-vessel classification; 0.8 reduction factor on allowables; acceptance of sub-15~psig relief devices. \\
            FESHM 5031.1               & Documentation thresholds and exemptions for cryogenic piping. \\
            FESHM 5032                 & Cryogenic safety review. \\
            FESHM 5033                 & Vacuum-vessel safety analysis; venting guidance referencing API~520 and CGA\,S-1.3. \\
            FESHM 5064                 & Oxygen-deficiency hazards; monitoring and alarms. \\
            \bottomrule
        \end{tabular}
    \end{table}

    The next section presents the design and analysis of the \textsuperscript{4}He evaporation refrigerator. The FNAL engineering notes are the primary documentation submitted to the CSP for FESHM compliance review and are therefore summarized in detail.

\section{Compliant Design}
\label{system_overview}

    The \textsuperscript{4}He evaporation refrigerator operates near 1~K and comprises three integrated subsystems: a vacuum-insulated shell-and-nose assembly that houses the evaporative stage, a cryogenic pumping stack that drives evaporative cooling, and a remote instrumentation and control architecture required by the radiation environment. Liquid helium is drawn from the superconducting magnet dewar reservoir through the U-Tube jumper into the phase separator and then delivered to the refrigerator nose.

    \begin{figure}[h!]
        \centering
        \includegraphics[width=\columnwidth]{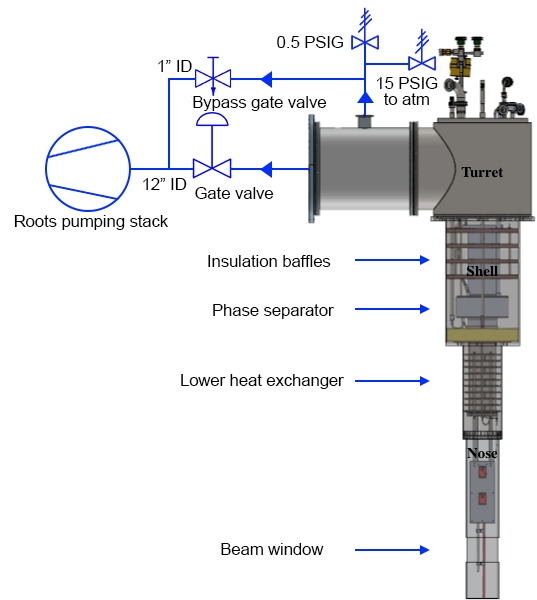}
        \caption{Simplified schematic of the refrigerator system showing the major components and connections to the pumping stack through the gate valve and bypass line. Arrows identify key features within the shell--nose assembly. The relief line connects to the main magnet relief, a 0.5~psig helium-recovery operational relief, and a 15~psig safety relief that vents to atmosphere.}
        \label{fig:block}
    \end{figure}

    The superconducting magnet provides the field required for DNP, stores the transfer liquid for the refrigerator, and insulates the underside and lateral portions of the refrigerator. The refrigerator shell and nose mate together and sit on the central axis of this insulated environment. A stainless-steel (SS) annular helium phase separator is located at the base of the shell and is thermally isolated from the warm upper exterior by a stack of perforated copper baffles. These insulation layers and upper baffles minimize the environmental heat load on the separator.

    Cold helium vapor is pumped out of the separator while liquid \textsuperscript{4}He is drawn in from the magnet. This process maintains a liquid reservoir in the separator at approximately 2.5~K. The separator then supplies liquid helium to the lower heat exchanger and to the nose reservoir, which is maintained as low as 1~K during normal operation. The target material is located in the beamline within the nose reservoir, where the 120~GeV proton beam enters through the beam window and deposits energy in the solid-state cryogenic target, NH\textsubscript{3} or ND\textsubscript{3} for SpinQuest (Fig.~\ref{fig:block}).

\begin{figure}[h!]
        \centering
        \includegraphics[width=\columnwidth]{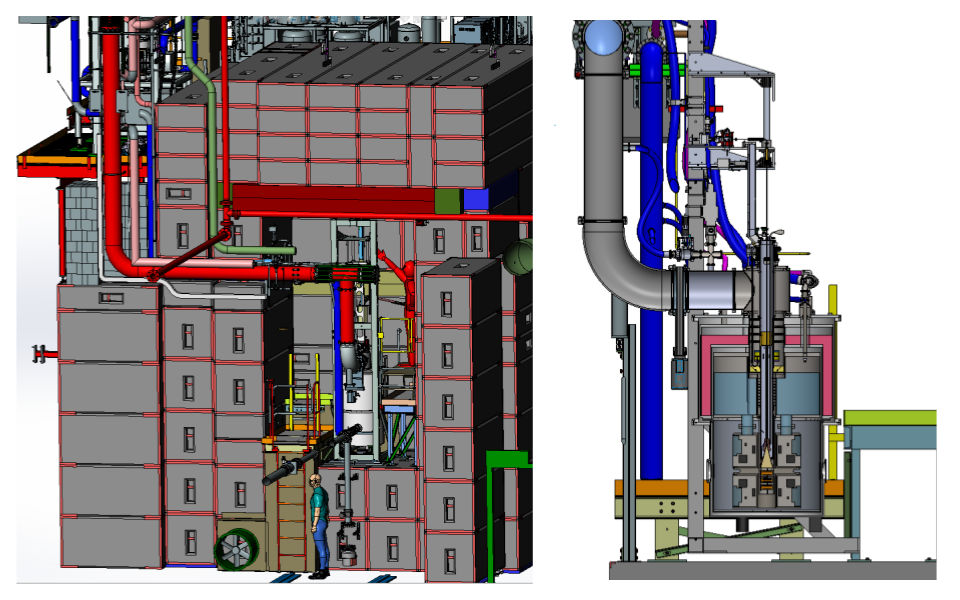}
        \caption{Target inside the radiation-shielded enclosure (left) and close-up of the refrigerator placement within the polarized target assembly (right).}
        \label{fig:targetcave}
    \end{figure}

    The 120~GeV, 10\textsuperscript{13}-proton-per-spill beam generates a radiation field that would damage downstream instrumentation and exceed the exposure limits established by the Fermilab radiation safety assessment. The polarized-target cryostat is therefore housed inside a dedicated radiation-shielded enclosure, the target cave, shown in Fig.~\ref{fig:targetcave}.

    The target enclosure is built from dense shielding blocks that absorb hadrons and attenuate secondary radiation from beam-target and beam-dump interactions, confining activation to the immediate interaction region. All silicon-based instrumentation is routed outside the enclosure. Within approximately 1~m of the target, electronics would experience mixed-field radiation of order 10\textsuperscript{3}~Gy(Si) total ionizing dose and 10\textsuperscript{12}~cm\textsuperscript{-2} (1~MeV neutron-equivalent) hadron fluence over a three-month run, sufficient to cause severe single-event effects and cumulative damage in most silicon devices within hours of beam operation \cite{semirad}. This condition motivates the remote-instrumentation strategy described in Section~\ref{Control}. Sensing lines connected to the refrigerator internals are routed through SS tubing, and the radiation-hardened control-valve design is presented in Section~\ref{Control}.

    Exhaust handling and cryogenic pumping are provided by a pumping stack consisting of Roots pumps for the refrigerator and a diaphragm pump for the separator. Overpressure scenarios in the exhaust path are handled by a relief network that sets the overall MAWP of the refrigerator. Relief sizing and stored-energy calculations are summarized below.

    The following subsections describe the main components of the \textsuperscript{4}He evaporation refrigerator, including geometric design, material selection, operational requirements, pressure analysis, hazard analysis, and relief sizing.

\subsection{Magnet Dewar}
\label{magdewar}
    The Oxford superconducting magnet supplies both the central cryogen reservoir for the refrigerator and the thermal insulation surrounding the magnet--refrigerator interface. The magnet helium space consists of two connected stainless-steel vessels with a maximum diameter of 27.5~in. The dewar can hold approximately 20~kg, or 135~L, of liquid helium. The superconducting solenoid is submerged in the lower magnet vessel, which connects to the larger upper vessel through two ports.

    Because the magnet vessel is classified as an experimental pressure vessel under FESHM Chapter~5031 \cite{fes}, a detailed engineering note was required to document pressure limits and hazard analysis, including the stored energy released during quench and instantaneous loss of insulating vacuum. A piping engineering note was also required under FESHM~5031.1 for each relief line \cite{fes}.

    Boiler Inspection Services Company performed nondestructive visual and ultrasonic-thickness (UT) examinations of the magnet \cite{inspector}, producing dimensioned material drawings for each section. The magnet is constructed from three discrete sections. The annular nitrogen dewar provides external thermal insulation above and along the outside walls of the helium dewar. The annular helium dewar surrounds the central refrigerator volume. The vacuum space is the sealed insulating region between these dewars and the shell--nose region that contains the refrigerator. The liquid-nitrogen insulation vessel holds 115~L. Under UG-23, maximum allowable stress values were obtained from ASME BPVC Section~II, Part~D, Subpart~1. The stress values used in the calculations include the required 0.8 reduction factor for experimental vessels and are listed in Table~\ref{tab:magnetmat}.

    \begin{table}[h!]
        \centering
        \caption{Stress values from ASME BPVC Section~II, Part~D, Subpart~1, reduced by 0.8 for experimental vessels.}
        \resizebox{\columnwidth}{!}{%
        \begin{tabular}{c|c|c|c}
            Component & Material & Stress (psia) & Stress Used (psia) \\
            \hline
            Insulation Vacuum & 304 Stainless Steel & 14200 & 11,360 \\
            Helium Dewar & 304 Stainless Steel & 14200 & 11,360 \\
            Nitrogen Dewar & 5456 Aluminum & 10700 & 8560 \\
        \end{tabular}
        }
        \label{tab:magnetmat}
    \end{table}

    The internal and external MAWP were determined for each vessel. Equation~\eqref{PIlongUG27}, from ASME UG-27 for cylindrical shells under longitudinal stress \cite{asme81}, gives the internal MAWP. Equation~\eqref{PeUG28}, from UG-28 for cylindrical shells and tubes with $D_o/t \geq 10$, gives the external MAWP.

    \begin{equation}
        P_i = \frac{2S E t}{R_o - 0.4t}
        \label{PIlongUG27}
    \end{equation}

     \begin{equation}
        P_e = \frac{4B}{3(D_o/t)}
        \label{PeUG28}
    \end{equation}

    A joint efficiency of $E=0.70$ was used for the vessel-shell calculations, consistent with ASME Section~VIII, Division~1, Table~UW-12 for double-welded butt joints without radiographic examination \cite{asme81,inspector}. The ASME factor $A$ for both materials was obtained from Fig.~G in ASME Section~II, Part~D, Subpart~3 \cite{asme2}. The ASME factor $B$ was obtained from Fig.~HA-1 and Fig.~NFA-10, respectively, in ASME Section~II, Part~D, Subpart~3 \cite{asme2,inspector}. The calculations used FNAL-provided stress values $S$ and the outer diameter $D_o$, total length $L$, outer radius $R_o$, and wall thickness $t$ from the UT inspection \cite{inspector}. The parameters and results from Eqs.~\eqref{PIlongUG27} and \eqref{PeUG28} are listed in Tables~\ref{tab:PIMag} and \ref{tab:PEMag}.

    \begin{table}[h!]
        \centering
        \caption{Internal MAWP of the magnet vessels.}
        \resizebox{\columnwidth}{!}{%
        \begin{tabular}{c|c|c|c}
            Vessel & $R_o$ (in.) & $t$ (in.) & $P_i$ (psia)\\
            \hline
            Outer Insulation Vacuum & 8.758 & 0.064 & 58.3 \\
            Inner Insulation Vacuum & 2.610 & 0.252 & 798.8 \\
            Outer Helium Dewar & 13.75 & 0.061 & 35.3 \\
            Inner Upper Helium Dewar & 4.75 & 0.125 & 211.5 \\
            Inner Lower Helium Dewar & 2.513 & 0.072 & 230.3 \\
            Outer Nitrogen Dewar & 17.08 & 0.212 & 74.8 \\
            Inner Upper Nitrogen Dewar & 4.74 & 0.115 & 146.8 \\
            Inner Lower Nitrogen Dewar & 14.46 & 0.175 & 72.9 \\
        \end{tabular}
        } \label{tab:PIMag}
    \end{table}

    \begin{table}[h!]
        \centering
        \caption{External MAWP of the magnet vessels.}
        \resizebox{\columnwidth}{!}{%
        \begin{tabular}{c|c|c|c|c|c|c}
            Vessel & $D_o$ (in.) & $L$ (in.) & $t$ (in.) & $A$ & $B$ & $P_e$ (psia)\\
            \hline
            Outer Insulation Vacuum & 17.515 & 8.075 & 0.064 & 0.0007 & 13,750 & 67.0 \\
            Inner Insulation Vacuum & 5.219 & 8.075 & 0.252 & 0.01 & 14,000 & 901.3 \\
            Outer Helium Dewar & 27.5 & 15.75 & 0.061 & 0.00025 & 3,500 & 10.4 \\
            Inner Upper Helium Dewar & 9.5 & 6.8 & 0.125 & 0.003 & 12,500 & 219.3 \\
            Inner Lower Helium Dewar & 5.025 & 8.95 & 0.072 & 0.00125 & 9,750 & 186.3 \\
            Outer Nitrogen Dewar & 34.15 & 4.4 & 0.212 & 0.001 & 8,750 & 141.5 \\
            Inner Upper Nitrogen Dewar & 9.48 & 4.4 & 0.115 & 0.0045 & 8,750 & 141.5 \\
            Inner Lower Nitrogen Dewar & 28.91 & 17.55 & 0.175 & 0.00125 & 6,500 & 52.5 \\
        \end{tabular}
        }
    \label{tab:PEMag}
    \end{table}

    Based on Tables~\ref{tab:PIMag} and \ref{tab:PEMag}, the magnet MAWP is 35.3~psia, set by overpressure within the helium space. The MAWP of the insulation-vacuum layer surrounding the cryogen spaces is limited by the nose and beam-pipe window rather than by the larger vessel parameters. During operation, the insulation vacuum is held at $\leq 10^{-7}$~Torr using turbomolecular pumps. The beam-pipe window is a 0.0055~in.-thick Ti-51 foil with a maximum allowable stress of 43,560~psia \cite{window}. Because the beam-pipe window helps define a vacuum space under FESHM~5033, a relief and engineering note were required to document stress analysis \cite{fes}. A relief device is required for vacuum-insulation spaces that contain cryogen inventory to protect against cryogen leakage into the insulation space \cite{fes}. A previous technical note established a window design pressure of 15~psid \cite{window}. To prevent the insulation vacuum from reaching this pressure, a standard atmospheric pop-off safety relief valve was installed on the external Oxford vacuum can.

    \subsubsection{Nitrogen Vent Relief}
    FESHM~5031 was applied to the LN\textsubscript{2} dewar vent-relief line. The credible failure mode is overpressure caused by a failed check valve on the 1.5~in. copper vent line to the outdoor exhaust. A local safety relief was therefore required. The vessel has four venting ports connected to the outside exhaust system in the target cave. For nitrogen-vessel pressure relief, we selected a pressure relief valve (PRV) with a set pressure $P_{\mathrm{set}}$ of 35~psig. Equation~\eqref{ReliefPressure} was used to determine the relieving pressure $P_r$ in accordance with API~520, Section~5.4.2.1.3, Table~2 \cite{API}.

    \begin{equation}
    P_r = 1.1P_{\mathrm{set}} + 14.7~\mathrm{psia}
    \label{ReliefPressure}
\end{equation}
    This gives a relieving pressure of $P_r = 53.2$~psia, which is within the limits in Table~\ref{tab:PIMag}. The calculation and vessel drawings were documented in two engineering notes \cite{ln2,ln22}.

    \subsubsection{Helium Exhaust Relief}
    The helium exhaust relief protects the magnet dewar against credible overpressure events such as quench or sudden loss of insulating vacuum. ASME UG-21 states that the design pressure should account for at least the most severe coincident pressure condition and may be taken as the difference between the outside and inside MAWP \cite{asme81}.

    The most critical credible overpressure event is simultaneous magnet quench and catastrophic loss of insulating vacuum. A relief analysis sized the helium relief system against this case \cite{quench}. Loss of vacuum was modeled using the Lehmann and Zahn experimental heat-flux data applied to the 2.525~m\textsuperscript{2} of liquid-wetted surface. Quench heat input was derived from transient thermal FEA of the superconducting coils and their 316LN stainless-steel former. With a stored magnetic energy of 0.199~MJ, the bounding 100\% deposition assumption gives an initial heat load of $\sim$19.6~kW decaying with a $\sim$20.3~s time constant. The FEA, which credits energy retained in the coils and former, reduces the helium-side peak to $\sim$5~kW. The two contributions combine to a peak heat load of 19.174~kW at 10~s, requiring as much as 1084~g/s of saturated-vapor relief flow. A transient ANSYS CFX simulation of the discharge path, using a custom NIST RefProp real-gas helium property file, Dittus--Boelter wall heat transfer, and quadratic pressure-drop coefficients for the flex hoses, piping, and burst disc, showed a peak pressure of 12~psig at the flex-hose inlet. This is below the 13.6~psid allowable drop set by the direct-integration orifice model of the risers and check valves. The vessel is therefore relieved below the MAWP using a design pressure and relief setting of 26.696~psia (12~psig + 1~atm) \cite{quench}, which keeps the vessel below the peak transient pressure that develops during the worst-case event.

    The relief line consists of three risers from the magnet that flow through three parallel 2~in. ID flex hoses, each 135~in. long. These lines connect to a common header with a single 4~in., 12~psig Oseco PLR burst disc that vents to atmosphere through a 35~ft-long, 4~in. Schedule~5 exhaust pipe. ASME UG-153 limits overpressure to the greater of 10\% of MAWP or 3~psig, giving a maximum required relief pressure of 29.696~psia.

    An additional operational quench line was constructed. This line is directly connected to the magnet safety-relief volume, and the rupture disc provides safety protection for both volumes. The line functions as an operational relief path using an FNAL-built 4~in. parallel-plate relief set at 5~psig with a capacity of 1681~g/s. The magnet safety relief is the main relief path and defines the MAWP of the refrigerator as a whole. With the 3~psig overpressure allowed by ASME, the maximum safety relief for the refrigerator is limited to 15~psig \cite{asme81}. The quench line vents to the helium recovery system, and the shell and nose are connected to this line so that all operational relief flow is routed for recovery.

    \subsection{Shell and Nose Assembly}
    \label{ShellNose}

    The shell and nose assembly provides the high-vacuum insulated barrier and structural support for the refrigerator internals, forming the refrigerator space. The shell is bolted to the top flange of the 5~T superconducting magnet, which serves as the inner wall of the insulating vacuum chamber. The turret connects the pump stack to the refrigerator space and is bolted to the top flange of the shell. The refrigerator is mounted on the turret top flange and aligned with the vertical axis of the magnet. The target insert passes through the central bore of the assembly and is sealed to the top of a bellows line with a Ladish flange and clamp; the bellows line connects to the top flange of the refrigerator. The full assembly is 46.7~in. long, with a top flange diameter of 12.6~in. tapering to 4.8~in. These dimensions allow the phase separator to be positioned along the vertical axis of the magnet bore without interference. The overall geometry is dictated by the 5~T superconducting magnet, to which the refrigerator cryostat must couple directly. The refrigerator channel measures 2.5~cm~$\times$~8~cm to accommodate the polarized target cell and insert.

    Material selection for the shell and nose prioritized fabrication, thermal performance, and code compliance. The shell was fabricated from rolled 304 stainless-steel sheets joined by fusion welds, and the nose was machined from a solid aluminum 6061-T6 rod. Qualified personnel performed the fusion welds in accordance with FESHM requirements \cite{fes}. The shell and nose were bolted together and sealed with an indium gasket to ensure vacuum integrity at low temperature. This design provided a practical assembly aligned with the experimental requirements while supporting FESHM compliance. The shell and nose assembly falls under the low-volume exemption (0.79~ft\textsuperscript{3}) from FESHM~5033 \cite{fes} and is directly connected to the magnet relief line. The CSP therefore did not require a formal engineering note. As a supplement, qualified personnel prepared an engineering document with drawings and pressure-limit calculations. ASME calculations were then performed to verify pressure compliance.

    Section~VIII UG-27 calculations were performed to determine the internal pressure limit \cite{asme81}. The limiting component is the thin aluminum 6061-T6 nose window, machined from solid rod, with weld efficiency $E=1$ from ASME Section~II, Part~D, Table~UW-12 \cite{asme81}. For a cylindrical shell under circumferential stress with radius $R = 1.87$~in. and thickness $t = 0.013$~in., the internal MAWP is given by Eq.~\eqref{eq:MAWP}.

    \begin{equation}
        P = \frac{S E t}{R + 0.6t}\label{eq:MAWP}
    \end{equation}
    Here $S$ is the maximum allowable stress specified in UG-23. Using the ASME Section~II, Part~D allowable stress $S = 12.0$~ksi \cite{asme81}, the calculated internal MAWP is 97.7~psia. As noted above, the refrigerator relief is limited by the magnet relief line to 15~psig, which sets the system MAWP. Using Eq.~\eqref{stress} at the design pressure $P_d$, the calculated stress is 2.1~ksi, a factor of five below the allowable stress.

    \begin{equation}
        \sigma = \frac{P_{d}R}{t}\label{stress}
    \end{equation}

    During operation, the refrigerator maintains the liquid \textsuperscript{4}He level and temperature during beam-target interactions. The beam-interaction region is defined by a precision 0.013~in.-thick beam window. This thin window minimizes beam scattering from the aluminum cryostat and defines the beam path through the nose so that the beam interacts predominantly with the polarized target material.

    FESHM~5033, Section~1.3, ``Safety Factor---Unmanned Areas,'' allows vacuum windows in unmanned areas to have minimum failure pressures approaching the external differential pressure. A credible failure mode would occur if the Roots pumps were turned on before the insulation vacuum was established. Therefore, the external MAWP was also evaluated using UG-28. Equation~\eqref{external} gives $P_a=8.22$~psia using the target-window outer diameter $D_0=3.77$~in., the modulus of elasticity $E=11.1\times10^{6}$~psia for aluminum 6061-T6 at $325^\circ$F from ASME Section~II, Part~D, Table~NM-2, and factor $A=0.00032$ from Fig.~G of ASME Section~II, Part~D, Subpart~3 \cite{asme2}.

    \begin{equation}
        P_{a}= \frac{2 A E}{3(D_{0}/t)}\label{external}
    \end{equation}

    A safety relief valve with a 0.5~psig set pressure was installed to keep the target window compliant. With this relief in place, the failure scenario was tested to verify safe operation. The FESHM~5033.1 definition of a vacuum window excludes beam-passage portions of vacuum vessels that are designed using ASME \cite{fes}. Under this exception, the nose did not require a separate vacuum-window engineering note.

    To reduce the thermal load, the external surfaces of the shell and nose, excluding the beam windows, are covered with multilayer insulation consisting of Mylar layers separated by webbed mesh spacers. This insulation primarily reduces radiative heat transfer while limiting direct thermal contact between layers.

    \subsection{U-Tube Jumper}
    \label{utube}

    The Cryofab U-Tube serves as a low-resistance, vacuum-insulated transfer path for liquid helium from the magnet dewar to the phase separator. It mates to the magnet dewar and phase separator through a compression fitting and precision brass coupling to ensure leak-tight operation. The U-Tube is constructed from 316L SS with an inner diameter of 0.188~in. and a wall thickness of 0.016~in.

    Under FESHM~5033, the vacuum jacket requires positive-pressure relief to ensure safety if helium enters the jacket during an overpressure event. FESHM~5031.1 also applies because the U-Tube is a transfer tube connecting the magnet dewar to the separator vessel. Two engineering notes were therefore required: one for overpressure of the internal flow path and one for overpressure of the vacuum jacket. The U-Tube has a Cryofab design pressure of 27~psid and was pneumatically tested at 30~psig under ASME UG-100 \cite{asme81}. The relief device is a Cryofab positive-pressure plunger-type relief orifice, also used as the vacuum pump-out port, with an ID of 0.37~in. A drawing documented the piping and dimensions, and pressure-drop and flow calculations were performed to verify quench compliance.

    During a quench with insulation loss, a small amount of vapor can move from the magnet through the U-Tube transfer line to the separator volume. This path is much more restrictive than the magnet relief path to the parallel-plate relief, reducing the possible mass flow through the tube under this failure mode. The U-Tube mass flow can approach 0.5~g/s, which was conservatively rounded to 2~kg/h. Using Eq.~\eqref{deltaP utube}, derived from the Darcy--Weisbach equation with minor losses \cite{crane}, this flow produces a pressure drop of 0.387~psig. The calculation used helium density at 100~K of $\rho_{100}=0.48~\mathrm{kg/m^{3}}$, conductance $K_{U\text{-}Tube}=2.5$, and flow velocity $V_{he}=66.7~\mathrm{m/s}$.

    \begin{equation}
        \Delta P_{U\text{-}Tube}= K_{U\text{-}Tube}\frac{\rho_{100}V_{he}^{2}}{2}
        \label{deltaP utube}
    \end{equation}

    The 316L SS U-Tube is relieved at both ends through the separator and magnet relief paths. The magnet safety relief was credited for U-Tube compliance, and the U-Tube was treated as existing piping under FESHM~5031.1.

    \subsection{Phase Separator and Piping}
    \label{baffle}

\begin{figure}[h!]
        \centering
        \includegraphics[width=\columnwidth]{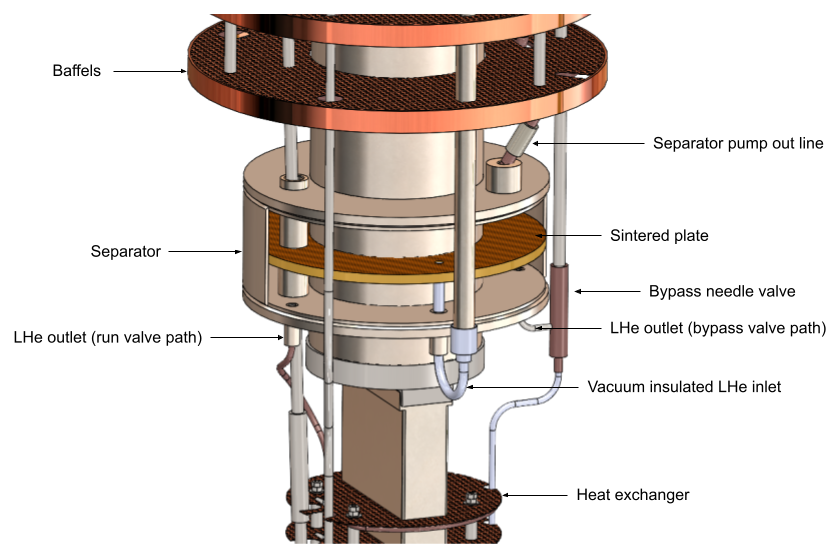}
        \caption{Cross-sectional view of the phase separator showing the baffles above and the liquid-helium path.}
        \label{fig:sintered}
    \end{figure}

    The annular phase separator, located at the base of the shell, provides the first stage of precooling and manages the liquid--vapor mixture transferred from the magnet dewar through the U-Tube. During operation, liquid helium transferred from the magnet dewar partially evaporates, forming a two-phase mixture. This flow passes from the U-Tube to a J-connection, a vacuum-insulated liquid-helium inlet that enters through the bottom of the separator and discharges onto the separator's middle sintered copper plate. The plate has 50~$\mu$m porosity and 1~mm thickness and divides the separator into two chambers. Manufactured from Electrolytic Tough Pitch (ETP) copper powder, the sintered plate provides high thermal conductivity and large surface area, enabling efficient vapor--liquid separation and evaporative precooling of the remaining liquid. The fabrication and design of the sintered plate are described in Refs.~\cite{Tapio,sinteredtapio}. A model of the separator with the sintered plate exposed is shown in Fig.~\ref{fig:sintered}.

    The separator has an outer diameter of 6.43~in. and an inner diameter of 6.37~in. Except for the sintered plate, the separator was fabricated from 304L SS with fusion welds so that it fits concentrically around the polarized-target insert at the base of the shell. All welds were performed by qualified personnel and inspected.

    As vapor and liquid separate, a diaphragm pump removes helium vapor from the separator and exhausts it upward across radial perforated copper baffles that are in thermal contact with the central bore of the refrigerator and the shell wall. Separator pumping simultaneously regulates the flow from the magnet to the separator and cools the top of the shell. The baffles act as a cold radiation shield, reducing environmental heat ingress. The precooled liquid from the separator flows down to the lower heat exchanger, where it enters the nose. This lower heat exchanger provides the second stage of precooling: liquid helium at $\sim$2.5~K from the phase separator is further cooled by $\sim$1~K vapor returning from the nose. Liquid flow into the exchanger is regulated by a PID-controlled needle valve, the run valve, while a parallel PID-controlled bypass needle valve allows rapid filling of the nose volume. These valves are discussed in Section~\ref{Control}.

    To improve heat transfer, the coiled heat-exchanger tubing is thermally anchored to several perforated copper baffles similar to the cold-insulation baffles. These baffles increase surface area, promote uniform cooling, and reduce heat leaks from the shell wall. The exchanger consists of 0.10~in.-diameter stainless-steel tubing extending 11.4~in. from the separator and coiled concentrically around the polarized-target insert. Stainless-to-stainless joints are fusion welded, while copper-to-stainless interfaces are joined with silver solder. This method is standard for rapid-cycling helium refrigerators and provides mechanical compliance for thermal contraction during repeated cooldown and warm-up cycles. The separator vessel forms a close fit that directs exhaust flow through the perforated copper baffles, strengthening the cold-insulation layer.

    The separator is both a cryogenic pressure vessel and a critical cooling component. It must withstand credible overpressure events, support robust control in a high-radiation environment, and provide the precooling required for efficient refrigerator operation.

    The separator vessel required an engineering note even though its volume was less than 35~ft\textsuperscript{3}, because the governing requirement is not a simple size cutoff. Under FESHM~5032, cryogenic vessels are reviewed according to type and code scope. For cryogenic pressure vessels, FESHM~5032 directs that FESHM~5031 applies whenever the vessel falls within ASME BPVC Section~VIII scope. FESHM~5031 then applies to any Fermilab vessel within that scope, and the scope determination must account for the full installation configuration and all possible pressure sources.

    The under-35-ft\textsuperscript{3} language in FESHM does not create a blanket exemption from engineering review. That exclusion appears in FESHM~5031.5 and FESHM~5033, which apply to specific vessel categories outside the primary ASME Section~VIII pathway. FESHM~5031.5 excludes certain low-pressure vessels under 35~ft\textsuperscript{3} from that chapter, but it explicitly states that the exclusion does not remove the vessel from Fermilab Engineering Manual processes. Similarly, FESHM~5033 excludes vacuum vessels under 35~ft\textsuperscript{3} only from the vacuum-vessel chapter; it is not a general waiver from pressure-vessel or cryogenic review.

    Because the separator is a cryogenic vessel, FESHM~5032 preserved review authority through the cryogenic process. It specifies that cryogenic pressure vessels within ASME Section~VIII scope must follow FESHM~5031, that low-pressure vessels outside ASME scope fall under FESHM~5031.5, and that other cryogenic vessels are reviewed as required by the Cryogenic Review Panel assigned to the system. It also clarifies that, in vacuum-insulated assemblies, the inner vessel is treated as a cryogenic pressure vessel and the outer shell as a vacuum vessel. The separator's small volume therefore did not eliminate the need for formal engineering documentation or ASME-compliant design, fabrication, certification, and stamping.

    Several conditions could cause separator overpressure. If the separator pump stops or the return valve closes, normal boil-off must be relieved. A second failure mode involves closing the helium return line while the magnet cryostat remains cold; if the return line is later opened, liquid helium could enter a warm separator and rapidly raise the pressure above 12~psig, stopping the transfer. During normal operation, the separator is in a helium atmosphere at 77~Torr. Loss of vacuum around the separator would impose a heat load on the vessel walls and create an overpressure event. A final scenario occurs during magnet quench, because the magnet and separator are connected through the U-Tube. Loss of vacuum in the refrigeration vessel was determined to be the most critical case because the U-Tube restricts flow to the separator during a magnet quench.

    Expansion of cryogenic fluid during loss of vacuum can produce higher-than-normal pressure in the refrigerator \cite{Pace}. Per FESHM~5031, a direct relief path to the vessel was implemented. This relief path, together with the upper and lower separator piping, required a piping engineering note under FESHM~5031.1.

    The helium separator vessel is a welded 304 SS vessel with wall thickness $\sim$0.065~in. It has a volume of approximately 52~in\textsuperscript{3} and a radius of 6.4~in. Code evaluation using the ASME Section~VIII thin-shell formula with weld efficiency $E=0.85$ gives a theoretical MAWP of 286.2~psig from Eq.~\eqref{eq:MAWP}. Following the same procedure as above, the allowable stress for 304L SS, $S=16.7$~ksi, is much greater than the maximum calculated stress of 5.7~ksi. With a design pressure of 40~psid \cite{sep-relief}, a 1/2~in. Circle Seal safety relief with a set pressure of 15~psig and a capacity of 81~g/s was approved, paired with a 3~psig operational relief. The operational relief valve provides a large margin against credible transients while maintaining compliance with FESHM guidance for sub-15~psig operation at the device level. This arrangement also complies with the maximum relief pressure set by the magnet relief line.

    Separator piping for the upper insulation baffles, lower heat exchanger, and dedicated relief path required a piping-specific engineering note under FESHM~5031.1 \cite{fes}. Detailed engineering analysis \cite{sep-relief} was performed for each path and is summarized below. The note verifies that the piping system remains safe outside its design pressure and temperature. Table~\ref{tab:piping} summarizes separator-connected runs and shows large margins relative to normal operating pressure.

    \begin{table}[t]
        \centering
        \caption{MAWP analysis for separator-associated piping and tubing.}
        \label{tab:piping}
        \begin{tabular}{@{}l l l l l@{}}
            \toprule Service        & Material & OD (in) & Wall (in) & MAWP (psig) \\
            \midrule Vent to Relief & 316L     & 0.50    & 0.035     & 2,279       \\
            U-Tube Fill             & 316L     & 0.188   & 0.016     & 2,872       \\
            Fill Line               & 304      & 0.435   & 0.020     & 1,620       \\
            Drain Line              & 304      & 0.065   & 0.0075    & 4,800       \\
            Lower HX                & Cu       & 0.125   & 0.0125    & 1,275       \\
            He Bypass               & 304      & 0.125   & 0.0175    & 6,222       \\
            Vent to Pumpout         & Cu       & 0.125   & 0.0125    & 1,275       \\
            Vent Header             & 304      & 0.865   & 0.0675    & 2,959       \\
            \bottomrule
        \end{tabular}
    \end{table}

    The adequacy of the relief-valve system was determined using the assumptions and calculations in Ref.~\cite{sep-relief}. The refrigerator was assumed to be operating or on standby at the start of a loss of insulating vacuum. In this event, the refrigerator shell warms and the liquid helium in the nose reservoir evaporates. The separator is surrounded by boil-off helium from the nose, which normally contains less than 2~L. It is conservative to assume an LN\textsubscript{2} temperature of 77~K for helium escaping from the separator because the relief path from the nose passes over the separator. The separator surface area is 0.0387~m\textsuperscript{2}.

    The maximum helium inventory in the separator is 0.096~kg, based on a normal operating pressure of 77~Torr. Air condensation onto the relief line was evaluated to verify that added latent heat would not cause failure, using Lehmann and Zahn data \cite{Lehmann} as detailed in the engineering note \cite{sep-relief}. The pseudo latent heat at a conservative 44~psia was interpolated as 17.119~J/g.

    The Lehmann and Zahn data were fit to Eq.~\eqref{eq:qlz}, which gives heat-flux density as a function of time. The integrated heat input and vented mass were then calculated using Eqs.~\eqref{qt} and \eqref{mass}.

    \begin{equation}
    \resizebox{\columnwidth}{!}{$
      \displaystyle
      \dot{q}_{LZ}(\tau_{LOV}) = \left[0.0097\left(\frac{\tau_{LOV}}{\mathrm{s}}\right)^3
      - 0.1363\left(\frac{\tau_{LOV}}{\mathrm{s}}\right)^2
      + 1.1802\left(\frac{\tau_{LOV}}{\mathrm{s}}\right) + 0.0262\right]\\,\frac{\mathrm{W}}{\mathrm{cm}^2}
    $}
    \label{eq:qlz}
    \end{equation}

    \begin{equation}
        q_t = \int_{t1}^{t2}(q_{LZ}(time))dt
        \label{qt}
    \end{equation}
\begin{equation}
    \Delta{M}_t =\frac{q*area_{outside}}{\Delta{h}_{pseudo}}
    \label{mass}
\end{equation}

    The venting rate was calculated using Eq.~\eqref{ventrate}.
\begin{equation}
    flow =\frac{q_{LZ}*area_{outside}}{\Delta{h}_{pseudo}}
    \label{ventrate}
\end{equation}

    The helium in the vent line was calculated to warm to 5.5~K during the first second, which was deemed negligible. This analysis supported approval of a 15~psig valve as the separator safety relief. FNAL personnel performed visual inspection and dimensional verification to confirm fabrication compliance.

    Thermal modeling of conductive heat transfer along the cryostat shell and nose was performed to estimate the environmental heat load without cold insulation. The model divided the system into three thermal regions: upper shell, lower shell, and nose, each corresponding to a distinct temperature regime.

    Under warm ambient conditions, the total conductive heat load is estimated to be 5.6~W. As the system cools to 100~K, the heat load decreases to approximately 1~W because of reduced thermal gradients and material conductivity. At 30~K, representative of operation with multilayer insulation in place, the heat load is further reduced to approximately 0.1~W.

    During operation, the helium separator maintains the lower shell region near 2.5~K, ensuring effective cold insulation and minimizing heat load. Pumping on the separator removes vapor and lowers the liquid temperature, establishing the sub-4~K environment required for the evaporative stage. The separator precooling stage improves thermal stability and cooling efficiency. The primary cooling power during target operation is provided by the high volumetric flow of the Roots pump stack, which lowers the pressure in the nose below 1~Torr and drives the nose liquid to 1~K.

    \subsection{Pumps and Roots-Pump Piping}

    Pumping on the helium bath lowers the pressure and induces rapid evaporation of liquid helium. This evaporative cooling removes heat from the polarized target sample and lowers its temperature to 1~K \cite{Tapio}. The pumping system therefore addresses the high-cooling-power requirement identified in Section~\ref{sec:barriers}.

    The pump stack consists of two WHU~7000 Roots blowers operating in parallel as the first stage (16,800~m\textsuperscript{3}/h at 60~Hz), a single WHU~7000 Roots blower as the second stage (8,400~m\textsuperscript{3}/h), and an SV630 rotary-vane backing pump (756~m\textsuperscript{3}/h). Conductance limitations in the Roots-pump piping reduce the ideal pumping speed. The resulting nominal pumping speed is 10,080~m\textsuperscript{3}/h, specified for inlet pressures up to approximately 0.1~mbar. Table~\ref{tab:pumps} summarizes the stack specifications.

    \begin{table}[h!]
        \centering
        \caption{Pump-stack configuration and capacities.}
        \resizebox{\columnwidth}{!}{%
        \begin{tabular}{l l l c }
            \hline
            Stage & Pump Type               & Model        & Capacity (m\textsuperscript{3}/h) \\
            \hline
            1     & Roots blower (parallel) & 2 $\times$ WHU 7000 & 16,800                       \\
            2     & Roots blower            & WHU 7000     & 8,400                        \\
            3     & Rotary vane             & SV630        & 756                          \\
            \hline
        \end{tabular}
        } \label{tab:pumps}
    \end{table}

    The Roots-pump piping connects the 1~K cryostat to a vertical intermediate reservoir and then to a large turret through a right-angle connection. This piping is also connected to the magnet relief path. FESHM~5031.1 required a piping engineering note and a drawing of the piping \cite{fes}.

    \subsection{Instrumentation and Remote Location}

    As established in Section~\ref{sec:barriers}, the target-cave radiation environment prevents long-term operation of silicon-based electronics near the cryostat. To minimize radiation exposure and facilitate maintenance, all active instrumentation, including pressure transducers and manometers, was placed outside the target cave approximately 20~m from the cryostat. Radiation-resistant elastomer or aluminum seals were used on cryogenic connections and vent piping, and a periodic replacement plan was established to ensure long-term reliability. Signals are transmitted through a stagnant 1/4~in. SS sensing line terminated with KF-16 fittings. The line is classified as Category~D service; when assembled from components with vendor ratings above 15~psig, it complies with FESHM~5031.1 \cite{fes} without requiring a formal engineering note \cite{manometer}. Under FESHM~5031.1, a formal engineering note is required if stored energy exceeds 150,000~ft$\cdot$lbf or if other exceptional conditions apply. The stored energy of the sensing line was evaluated \cite{manometer}.

    At room-temperature conditions, the Baker equation \cite{osti_967234} gives the vessel burst energy $W_{RT}$ in ft-lbf as

    \begin{equation}
        W_{RT} = \frac{P_1 Vol_{r}}{k_{RT} - 1}\left[1 - \left(\frac{P_1}{P_2}\right)^{\frac{1-k_{RT}}{k_{RT}}}\right] = 1.305\times10^3 .
        \label{burst}
    \end{equation}
    The calculation used initial pressure $P_1=30$~psig, ambient pressure $P_2=14.5$~psig, refrigerator volume $Vol_r=0.79$~ft\textsuperscript{3}, and specific-heat ratio $k_{RT}=1.666$. The result is well below the 150,000~ft$\cdot$lbf limit, justifying the decision not to require a piping engineering note \cite{manometer}.

    Refrigerator leak checks follow Fermilab's low-stress piping practice and are scheduled periodically. Although most instruments are located outside the target cave, some components, such as target-insert temperature sensors, must be placed directly in the radiation environment. These sensors are replaced at regular intervals to manage radiation-induced degradation.

    The radiation-hardened control approach is described in Section~\ref{Control}. Several sensors are held within the refrigerator helium space through the target insert.

    \subsection{Target Insert}
\begin{figure}[h!]
        \centering
        \includegraphics[width=8cm]{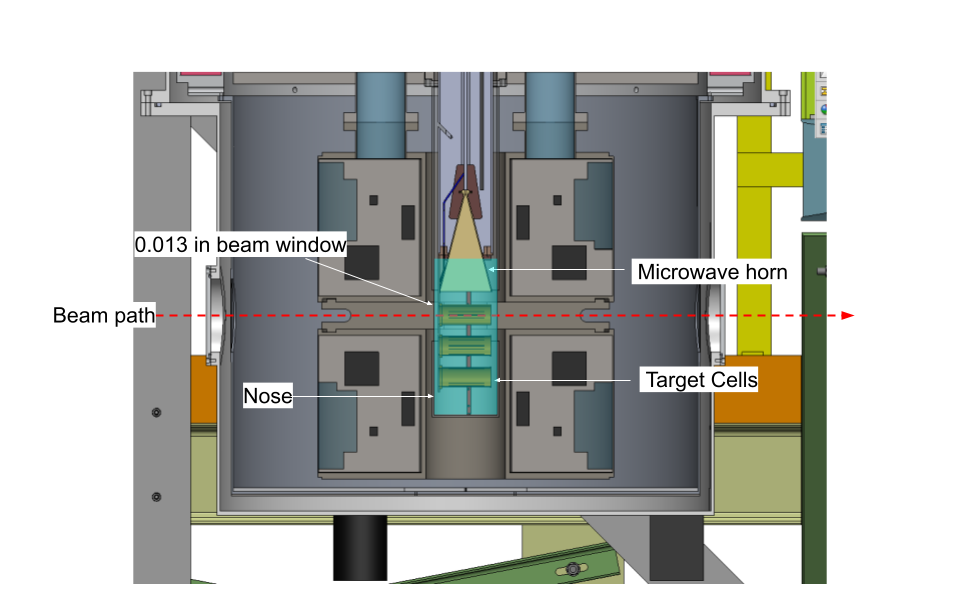}
        \caption{Close-up model of the target housing along the central axis of the refrigerator. The microwave horn delivers 140~GHz radiation to the target cells. Three target cells are shown with NMR loops inside. The beamline, shown as a red dashed line, passes through the beam window and interacts with the target. All three target cells are submerged in liquid helium.}
        \label{fig:stick}
    \end{figure}

    The target insert, shown in Fig.~\ref{fig:stick}, is placed along the central axis of the refrigerator and contains three cells for target material. Each elliptical cell measures 20~mm~$\times$~27~mm, is 8~cm long, and is made of polychlorotrifluoroethylene (PCTFE), selected for its low-temperature properties and radiation hardness. Each cell has three NMR inductor coils for polarization monitoring and three chip resistors for mapping the microwave profile along its length through temperature changes. Cernox sensors are placed on the top and bottom cells to provide secondary temperature measurements of the target material, complementing the helium vapor-pressure measurement. Accurate temperature monitoring is critical for optimizing annealing, which removes radiation-induced unpaired electrons accumulated during operation.

    The insert has a total length of 64.5~in., with a carbon-fiber body to minimize thermal conductivity. The top flange is a 4~in. Ladish blank flange containing SMA connectors for NMR, a microwave waveguide connection, and connectors for the chip resistors and Cernox sensors. Microwaves used for DNP are coupled from the microwave source through the waveguide interface at the top flange. The waveguide runs along the insert and terminates in a gold-plated horn positioned just above the target-cell assembly. This horn launches the microwave field into the liquid-helium volume and onto the target, providing efficient coupling and spatial coverage for all three cells.

    Figure~\ref{fig:stick} shows a close-up of the target housing, including the target cells, microwave horn, refrigerator nose, 0.013~in. beam window, and beam path.

    The insert enables stable DNP operation and continuous polarization monitoring across all three target cells under the radiation and cryogenic conditions of the SpinQuest target cave. It completes the design overview and system specification.

    Table~\ref{tab:MAWP} summarizes MAWP and normal operating pressure (NOP) for pressure-relevant refrigerator components. For example, the refrigerator bellows have a burst strength of 104~psig while operating at $\sim$14.7~psia. These values confirm that normal operation remains well within code-prescribed margins.

    \begin{table}[h]
    \centering
    \caption{Maximum allowable working pressure (MAWP) and normal operating pressure (NOP) for refrigerator components.}
    \label{tab:MAWP}
    \resizebox{\columnwidth}{!}{
    \begin{tabular}{@{}llll@{}}
    \toprule Component & MAWP (psia) & NOP (psia) & Source \\ [0.5ex]
    \midrule Refrigerator Bellows & 118.7 & 13.8 & Vendor (KeyHigh) \\
    U-Tube & 2872 & 13.8 & Cryofab \\
    Separator & 300.8 & 13.8 & In-house analysis \\
    Nose Window & 97.7 & $10^{-4}$ & ASME VIII calc. \\
    Insulating Vacuum & 29.7 & $10^{-8}$ & Oxford \\
    \bottomrule
    \end{tabular}
    }
    \end{table}

    Table~\ref{tab:system_summary} summarizes the system design requirements and safety specifications. These elements address the first key barrier: establishing a traceable, code-compliant design basis for a low-pressure cryogenic refrigerator.

    \begin{table*}[t]
        \caption{Summary of key SpinQuest \textsuperscript{4}He evaporation refrigerator specifications.}
        \centering
        \label{tab:system_summary}
        \begin{tabular}{l l l l}
            \toprule \textbf{Component} & \textbf{Key Parameters} & \textbf{Operating Condition} & \textbf{Safety/Relief} \\
            \midrule Shell + Nose       & Vol: 0.79~ft\textsuperscript{3}, 304SS + Al 6061-T6 & Vacuum: $10^{-3}$~Torr & FESHM~5033 exempt \\
            Phase Separator             & OD/ID: 6.4~in./3.6~in., 316L SS & Temp: sub-4~K, P: 3~psig & 3~psig relief valve \\
            U-Tube Jumper               & ID: 0.188~in., 316L SS & $\Delta P$: $<0.4$~psig & MAWP: 2872~psig \\
            Pump Stack                  & Capacity: 10,080~m\textsuperscript{3}/h & $P_{\mathrm{nose}}$: 0.1~mbar & N/A \\
            Instrumentation             & 20~m remote sensing lines & Category~D service & FESHM~5031.1 compliant \\
            \bottomrule
        \end{tabular}
    \end{table*}

\section{Measurement and Control}
\label{Control}

    Stable 1~K operation in the refrigerator nose is required for two central tasks: thermal-equilibrium (TE) polarization calibration and enhanced polarization of the target material \cite{Spin}. The absolute polarization scale is obtained by measuring the NMR response at thermal equilibrium and using that measurement to determine the calibration constant. Both TE calibration and high-polarization operation require precise control of helium level and temperature. The system provides this control through three remotely actuated valves governed by a PID-based LabVIEW architecture. This section describes the measurement basis and the control hardware and software that enable operation under the radiation and access constraints of the SpinQuest target cave.

    \subsection{Polarization}

    Target polarization is the principal operating parameter of the system. Crabb-style polarized targets use 140~GHz microwave sources applied to irradiated solid-state target materials in a 5~T holding field at 1~K. Refrigerator control is critical because the nose must remain filled with enough helium to keep the target submerged. A tuned NMR coil and Q-meter measure the polarization \cite{Spin}. The microwave RF field irradiates the spin system, which absorbs or emits energy \cite{Crabb,Tapio}. The response is described by the magnetic susceptibility $\chi$, with dispersive part $\chi'$ and absorptive part $\chi''$. The polarization is proportional to the absorptive component through Eq.~\eqref{Pol}.

    \begin{equation}
        P = K\int{\chi^{''}(\omega)d\omega}
        \label{Pol}
    \end{equation}

    Here $K$ depends on the efficiency of the NMR system. A tuned $\lambda$/2 coil antenna detects the response. Because the NMR signal is proportional to polarization but not absolutely calibrated, an independent reference measurement at known conditions is required. We calibrate the polarization using the calculated thermal-equilibrium polarization $P_{TE}$ at 1--2~K and compare this value with the enhanced-polarization signal to obtain the absolute polarization \cite{Crabb}. The accuracy of this calibration depends directly on the stability and accuracy of the temperature measurement during TE operation, as discussed in Section~\ref{TE}.

    \subsection{Thermal-Equilibrium\hspace{.1cm}Measurement}
    \label{TE}

    Thermal-equilibrium (TE) calibration establishes the absolute scale for NMR-based polarization measurements. At thermal equilibrium, the target polarization \(P_{\rm TE}\) follows the Boltzmann distribution for the populations of magnetic sublevels in a holding field \(B\) at temperature \(T\). For spin-\(1/2\) systems (e.g., protons in \(\mathrm{NH_3}\)) and spin-\(1\) systems (e.g., deuterons in \(\mathrm{ND_3}\)), the polarizations are given by
    
    \begin{equation}
    P(1/2) = \tanh\left(\frac{\gamma \hbar B}{2 k_B T}\right) \label{spin1/2}
    \end{equation}
    
    and
    
    \begin{equation}
    P(1) = \frac{4 \tanh\left(\frac{\gamma \hbar B}{2 k_B T}\right)}{3 + \tanh^2\left(\frac{\gamma \hbar B}{2 k_B T}\right)}, \label{spin1}
    \end{equation}
    
    respectively. Here \(\gamma\) is the gyromagnetic ratio of the nucleus, \(\hbar\) is the reduced Planck constant, and \(k_B\) is the Boltzmann constant.
    
    The absolute polarization is obtained from the ratio of the enhanced NMR signal (under DNP) to the TE signal, using the calculated \(P_{\rm TE}\) as the reference. This approach requires a precisely known temperature and magnetic field during the TE measurement.
    
    TE calibrations are performed with the target material fully submerged in a stable 1--2.5\,K liquid-helium bath, in the absence of microwave irradiation and beam-induced heating. These conditions eliminate external heat loads and ensure the target material reaches thermal equilibrium with the surrounding helium. Achieving and maintaining such stability is the primary motivation for the fine temperature control provided by the refrigerator. Helium vapor pressure (and thus bath temperature) is regulated via the remotely actuated gate-valve bypass, which controls the effective pumping speed on the 1\,K stage. Liquid level in the nose is independently maintained using the run and bypass needle valves. The design and remote actuation of these three control valves are described in the following subsection.
    
    \subsection{Control Valve Actuation}
\begin{figure}
        \centering
        \includegraphics[width=\columnwidth]{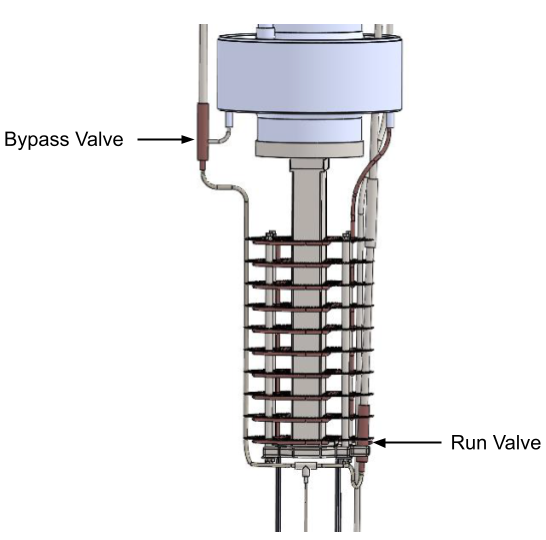}
        \caption{Run and bypass helium routes. The run and bypass needle valves are remotely controlled. The bypass valve reroutes liquid from the separator to the nose for fast filling. The run valve throttles precooled liquid from the separator through the lower heat exchanger and into the nose.}
        \label{fig:frige_valves}
    \end{figure}

    Because of radiation-induced failure risks and operational hazards, all control valves require remote operation. A model and hazard analysis were required for ORC approval before valve implementation and operation. The refrigerator uses three remotely operated control valves. Liquid flow into the exchanger is regulated by a PID-controlled Oxford needle valve, the run valve, while a parallel PID-controlled Oxford bypass needle valve allows rapid filling of the nose volume. Both are shown in Fig.~\ref{fig:frige_valves}.

    The primary design challenges were avoiding refrigerator overfill, which could damage brittle magnet vacuum seals; avoiding spatial interference with the target ladder and insert path; and preventing over-tightening of valves, which could deform equipment. To address these challenges, a NEMA~17 stepper motor with a friction clutch and flexible coupler was selected to regulate torque, with a 0.4~N\,m closing load. A custom SS spur gearbox with a 1.25:1 ratio offsets the motor assembly from the ladder path. Absolute position feedback is provided by a rotary potentiometer selected for radiation tolerance.

\begin{figure}
        \centering
        \includegraphics[width=\columnwidth]{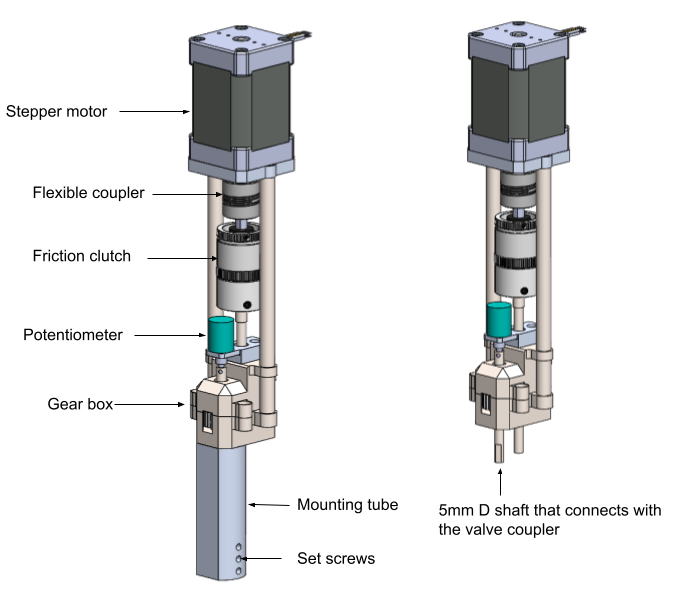}
        \caption{Run-valve assembly with custom gearbox, friction clutch, and rotary potentiometer. The design offsets the motor from the vertically moving target ladder and mounts to the run-valve shaft using three 8-32 set screws. A rotary potentiometer provides absolute valve-position feedback. The right image shows the 5~mm D-shaft.}
        \label{fig:runValve}
    \end{figure}

    The bypass valve required a horizontal layout to preserve target-insert clearance. A 10:1 worm drive was therefore used to redirect motion and was coupled through a friction clutch and potentiometer. Figure~\ref{fig:runValve} shows the run-valve assembly; the bypass-valve design is similar.

\begin{figure}[h!]
        \centering
        \includegraphics[width=5cm]{
            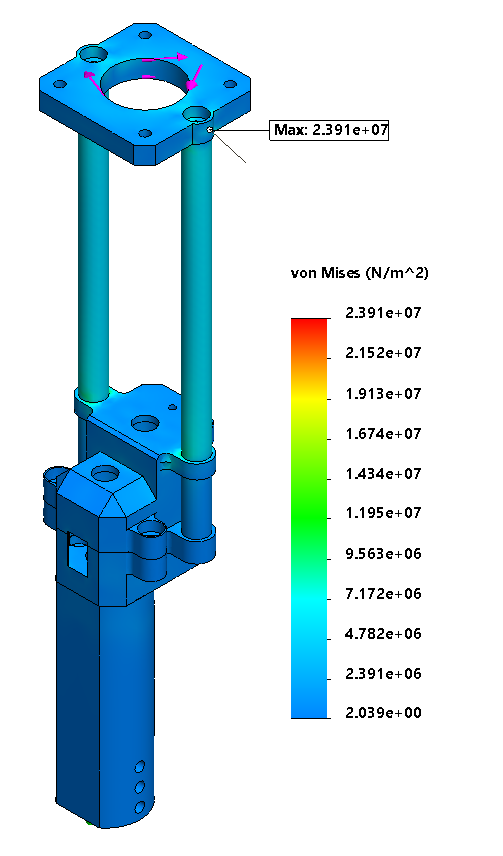
        }
        \caption{Maximum von Mises stress and displacement of the run valve under the worst-case loading scenario, simulated in SolidWorks for 304 stainless steel.}
        \label{fig:runV_sim}
    \end{figure}

    Both actuators were simulated at maximum motor torque without clutch slip to evaluate mechanical stability under extreme loading and to validate the design safety margin shown in Fig.~\ref{fig:runV_sim}. Feedback components were then tested to confirm radiation tolerance.

\begin{figure}
        \centering
        \includegraphics[width=\columnwidth]{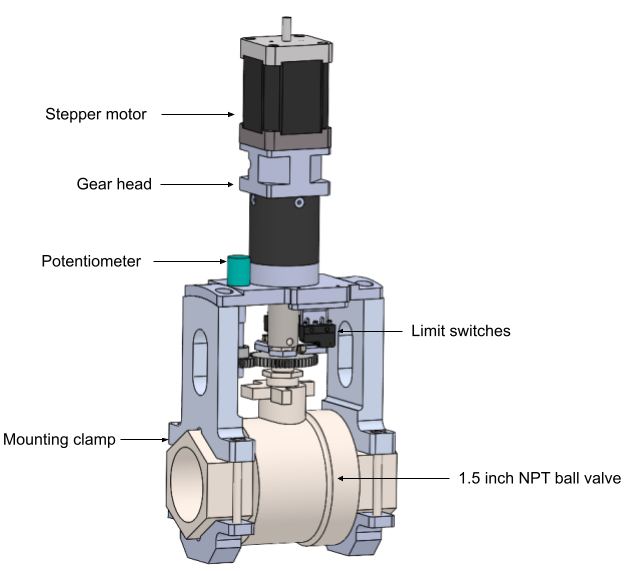}
        \caption{Gate-valve bypass actuator with 1:32 gearbox, stepper motor, rotary potentiometer for position feedback, and two travel-limit switches.}
        \label{fig:GVBypass}
    \end{figure}

    The third control valve, the gate-valve bypass, regulates helium vapor pumping speed during TE measurements, where fine temperature control is required. The refrigerator connects to the Roots pumping stack through two parallel paths: a 12~in. low-impedance gate valve for high-speed pumping and a 1~in. high-impedance motorized bypass line for precise flow control. The bypass actuator uses a stepper motor with a 1:32 planetary gearhead, potentiometer position feedback, and snap-action travel-limit switches, all selected for compatibility with the radiation environment and remote operation over a 20~m distance. A model of the remote gate-valve bypass is shown in Fig.~\ref{fig:GVBypass}.

During a TE measurement, the gate valve is closed, transferring full flow control to the gate valve bypass line. Once the system has reached thermal equilibrium in the absence of beam-induced heating, the TE measurement is performed by integrating the area of the NMR signal at a fixed target temperature. Stable target temperature, followed by stabilization of the NMR signal area, is required to obtain a high-quality measurement. TE measurements at lower temperatures take significantly longer due to the correspondingly longer $T_1$ nuclear spin-lattice relaxation times. The resulting TE signal area is used to determine the calibration constant required for absolute polarization determination.

    \subsection{Control System Architecture}

\begin{figure*}[h!]
        \centering
        \includegraphics[width=\textwidth]{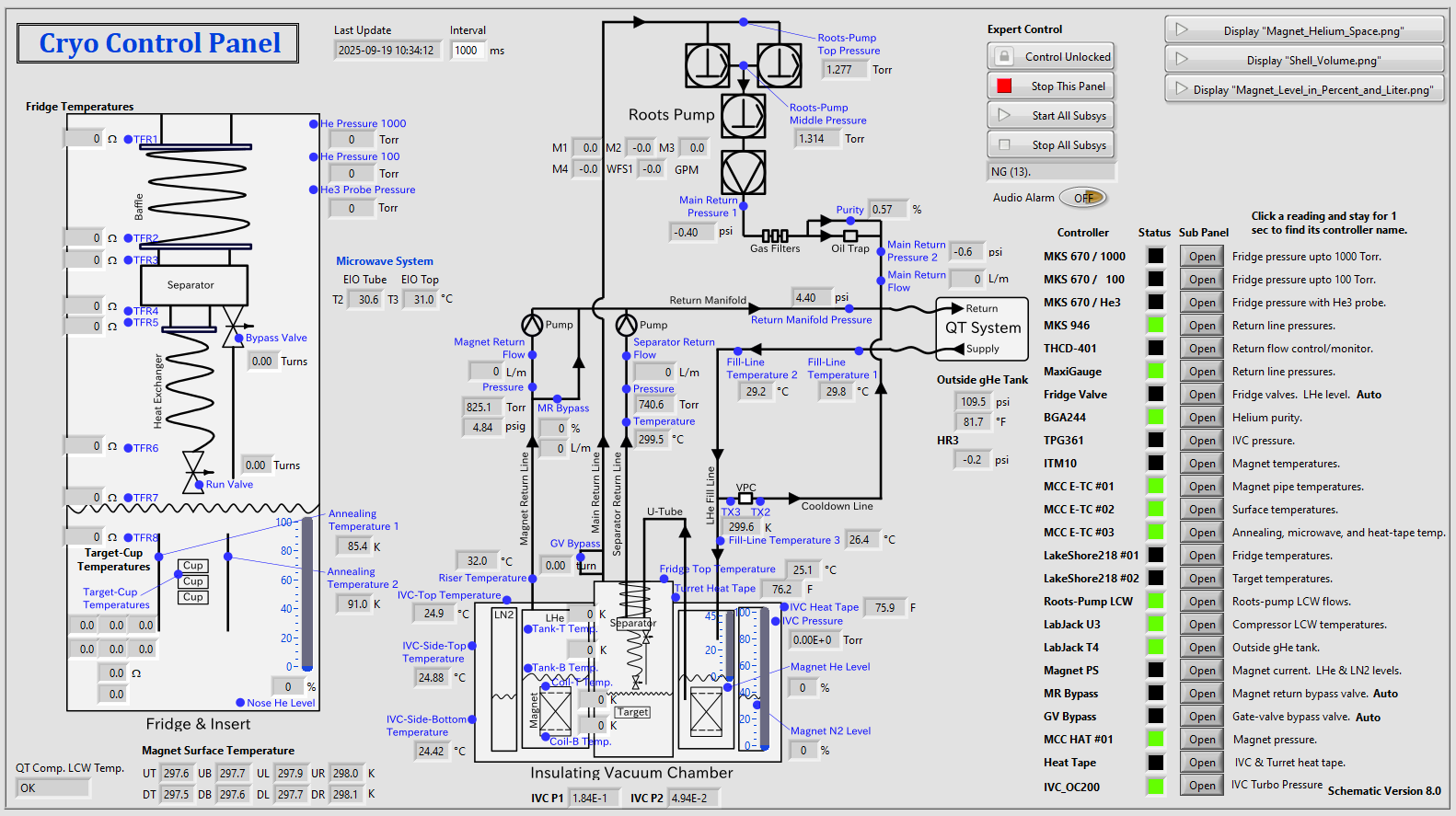}
        \caption{LabVIEW-based cryogenic control panel and live display for the SpinQuest refrigerator. The top-level VI supervises device-specific subVIs for valve actuation, pump and flow control, pressure and temperature readout, liquid-level monitoring, data parsing, alarm handling, and operator interlocks. The schematic displays the helium fill and return paths, Roots-pump system, separator and magnet return circuits, insulating vacuum chamber, refrigerator insert, control valves, and sensor locations. Manual and PID-based operation are available for the gate-valve bypass, run valve, bypass valve, separator pump, and magnet pump, with configuration parameters established during commissioning.}
        \label{fig:cryopannel}
    \end{figure*}

    Figure~\ref{fig:cryopannel} shows the LabVIEW-based cryogenic control panel for the SpinQuest main refrigerator and cryogenic system, excluding the liquefaction subsystem. The front panel is organized as a top-level, or mother, VI that supervises device-specific subVIs for valves, pumps, pressure transducers, flow meters, temperature sensors, liquid-level readout, data parsing, monitoring, and alarm handling. The central schematic provides a live piping-and-instrumentation-style representation of the cryogenic system, including the helium fill line, return manifold, Roots-pump train, gas filters and oil trap, separator and magnet return circuits, insulating vacuum chamber, refrigerator insert, and associated valves and sensors.

    Active control channels can be operated manually or through software PID loops during refrigerator operation. The gate-valve bypass can be configured for either flow- or pressure-regulated PID control. The separator return pump and magnet return pump each include flow-control loops that allow the operator to prescribe return-flow setpoints. The run valve is controlled using the nose helium-level probe as the liquid-level sensor, while the bypass valve is operated by a software liquid-level PID loop using the appropriate helium-level readback. Sensor instrumentation subVIs handle calibration, data parsing, display, status monitoring, and alarm generation. Channel assignments, PID gains, limits, and interlock thresholds were configured during commissioning.

    To prevent overfilling with liquid helium, the control VI automatically closes the run and bypass valves if the active filling time exceeds a preset limit, 10~min by default, or if the liquid level reaches the top of the heat exchanger. In addition to the nose helium-level probe, eight chip-resistor level indicators, TFR1--TFR8, are mounted along the refrigerator and used to track liquid level after it rises above the nose probe, covering the lower heat exchanger, separator, and upper baffle regions.

    For the run valve, helium flow and liquid level are the process variables. Flow is measured by mass-flow controllers on the main flow outlet, and liquid level is determined by a level probe. The bypass valve uses liquid level as its process variable. During TE measurements, the gate-valve bypass uses temperature as its primary process variable, inferred from helium vapor pressure measured by a Baratron manometer and cross-referenced with NIST \textsuperscript{4}He vapor-pressure--temperature correlations. This configuration enables closed-loop temperature stabilization between 1.3 and 4~K.

    The electronic control systems for each actuator are housed in separate 2U enclosures installed outside the target cave. Each enclosure includes stepper-motor drivers (Applied Motion ST5-S), a 5~V supply for potentiometers, and an MCC USB-202 ADC for position readout. The enclosures required FNAL ORC review to verify compliance with laboratory electrical-safety standards. Because the control room and target area are separated by 20~m, RS232 and USB interfaces were extended with USB-to-Ethernet converters.

    Remote control-valve operation, combined with PID control, maintains temperature stability during TE measurements. This instrumentation arrangement enables continuous helium-system monitoring while protecting sensitive electronics from the high-radiation environment inside the target cave. PHP-based web interfaces are used for online monitoring of measurement stability during operation.

    With the measurement and control architecture established, the second key barrier is addressed. The refrigerator performance under the three-barrier framework is evaluated in Section~\ref{commissioning}.

\section{Commissioning Performance}
\label{commissioning}

    The refrigerator was commissioned in July~2024 to evaluate environmental heat load, transient heat loads from the beam and DNP microwaves, cooling capacity, and operational stability. 
    The heat load of the system was estimated from the helium boil-off to give lower limits of the cooling power using the following equation,
    \begin{equation}
        \dot{Q} = \dot{n}\,L_{0},
        \label{cooling_power}
    \end{equation}
    where $\dot{n}$ is the helium mass flow rate and $L_{0}$ is the specific latent heat of vaporization, interpolated between 1~K and 1.5~K from NIST data~\cite{nistL}. Volumetric flow measurements were converted to mass flow using the helium density at standard conditions, $\rho_{\text{STP}} = 0.179~\mathrm{g/L}$.

    \subsection{Environmental Heat Load}
    \label{sec:env_heatload}

\begin{figure}[h!]
        \centering
        \includegraphics[width=\columnwidth]{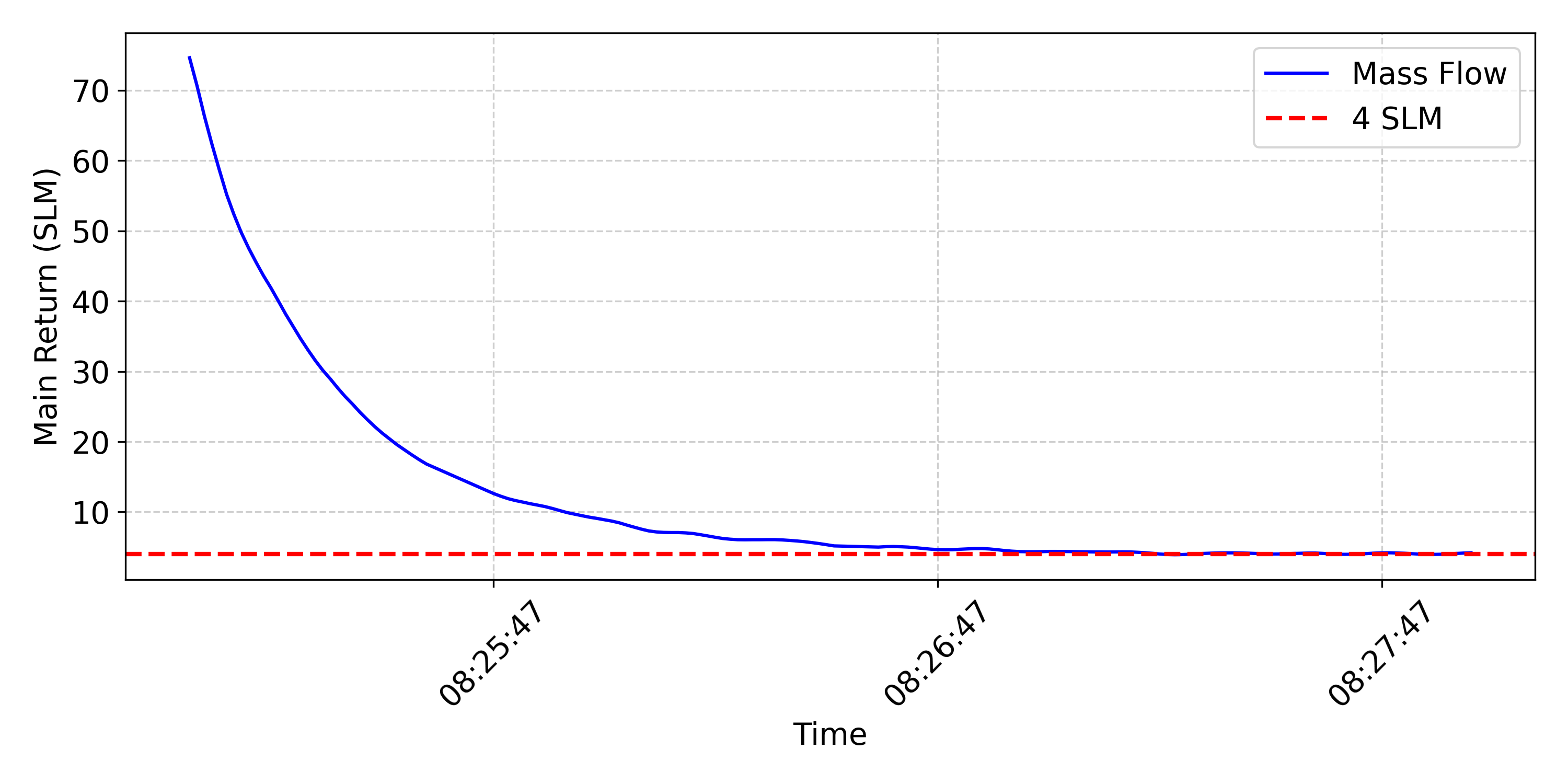}
        \caption{Steady-state boil-off flow following cooldown from atmospheric pressure with the run and bypass valves closed. The main flow trends toward $\sim$4~SLM at 1.36~Torr, corresponding to an environmental heat load of $\sim$0.2~W.}
        \label{fig:5slm}
    \end{figure}

    To establish a baseline environmental heat load, the run and bypass valves were closed and the system was allowed to reach thermal equilibrium below 2~K. The main flow stabilized at 4~SLM and 1.36~Torr (Fig.~\ref{fig:5slm}), corresponding to an environmental heat load of $\sim$0.2~W from Eq.~\eqref{cooling_power}. This value is consistent with the thermal-model prediction of 0.1--1~W developed in Section~\ref{baffle}.

    \subsection{Beam-Induced Heat Load}
    \label{sec:beam_heatload}

\begin{figure}[h!]
        \centering
        \includegraphics[width=\columnwidth]{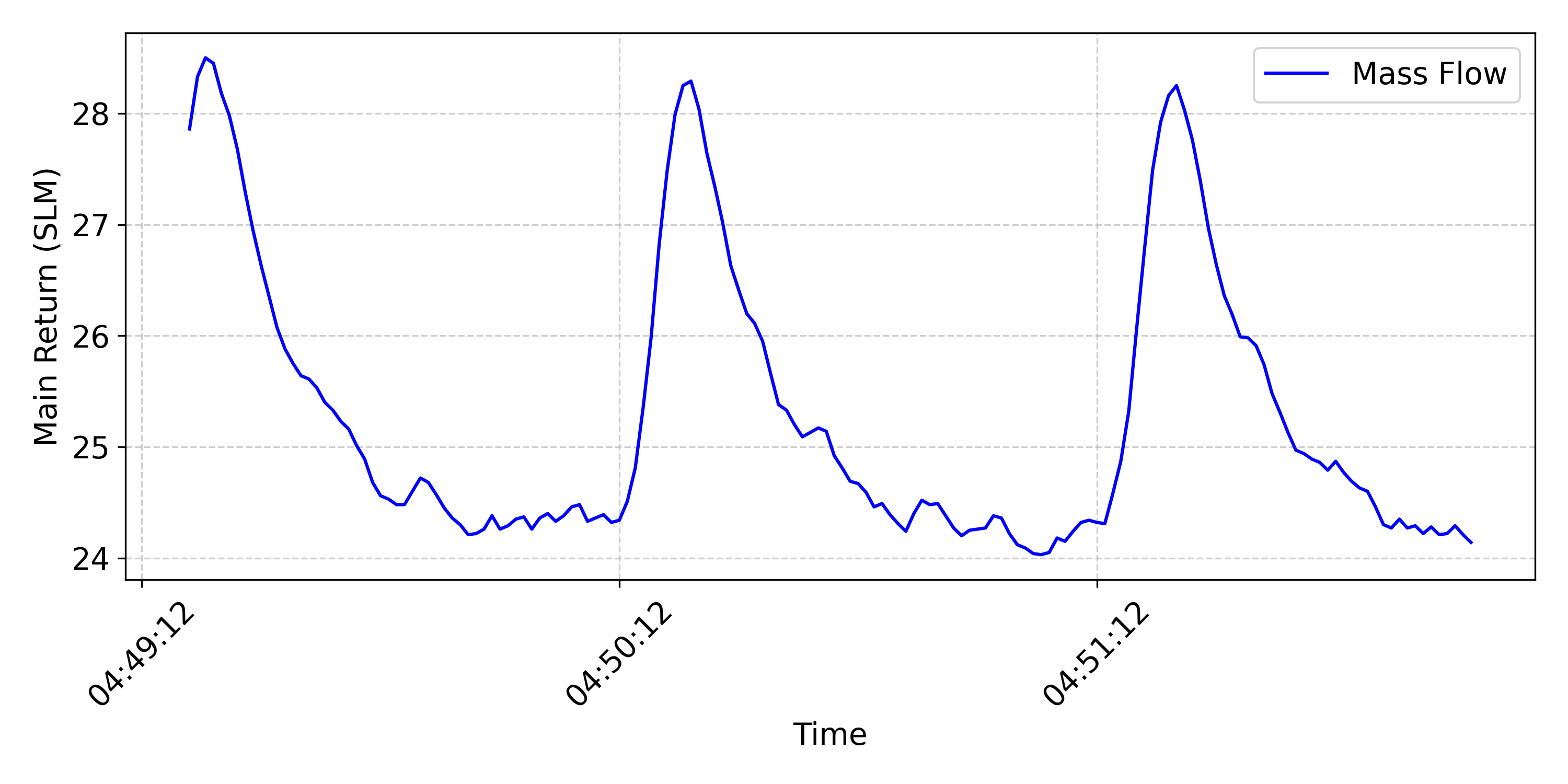}
        \caption{Main-flow stability under high-intensity beam operation. A flow increase of $\sim$4~SLM is observed once per 60~s beam cycle, corresponding to a beam-induced heat load of $\sim$0.2~W.}
        \label{fig:mainbeam}
    \end{figure}

    Beam-induced heat loads were characterized during high-intensity beam operation, with $10^{13}$ protons per 4~s spill at 120~GeV delivered to the SpinQuest target cave on a 60~s cycle. The refrigerator and cryostat maintained stable flow throughout data taking, with a periodic flow increase synchronized to each beam spill (Fig.~\ref{fig:mainbeam}). Each spill produced a $\sim$4~SLM increase above baseline flow, corresponding to a beam-induced heat load of an additional 0.2~W.

    \subsection{Thermal-Equilibrium Measurements via Radiation-Hardened Bypass Control}
    \label{sec:te_measurement}

Stable closed-loop pressure control achieved through the radiation-hardened gate-valve bypass is a key commissioning result. TE calibration of the target polarization using the NMR system requires the target temperature, regulated through helium vapor pressure, to remain constant over extended periods, ranging from tens of minutes to several hours depending on the target type. The gate valve bypass enables precise regulation of the high flow from the roots backing pump while the gate valve is closed and can be remotely actuated from outside the high-radiation target cave. This calibration is performed in the absence of microwave or beam-induced heat loads, thereby eliminating external heating sources and allowing the target material to reach true thermal equilibrium with the surrounding liquid helium bath.

\begin{figure}[h!]
        \centering
        \includegraphics[width=\columnwidth]{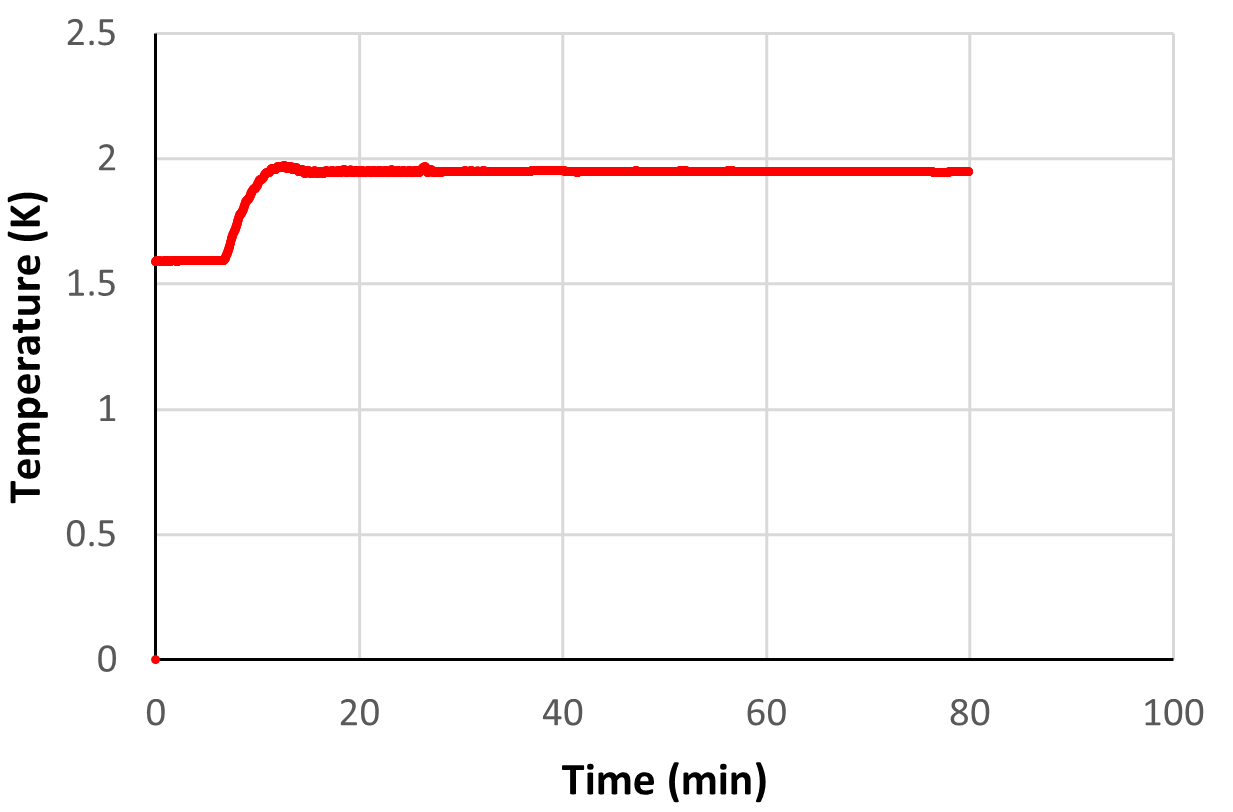}
        \caption{Temperature profile during a thermal-equilibrium calibration run on a CH\textsubscript{2} target. The NMR area used for polarization calibration began stabilizing within $\sim$20~min (driven by the material longitudinal relaxation rate $T_1$) under PID control of the radiation-hardened gate-valve bypass actuator.}
        \label{fig:p and temp}
    \end{figure}

CH\textsubscript{2} was selected for the initial TE calibration runs because it is operationally simpler than NH\textsubscript{3}, which was used during later commissioning and is the preferred polarized-proton target material for production data taking. A CH\textsubscript{2} target insert was filled and loaded into the cryostat. During these tests, the Roots boosters were disabled, only the backing pump was operated, and the outflow was regulated by the gate-valve bypass actuator under PID control while the main gate valve remained closed. The system was held at 20~Torr, and the NMR signal area stabilized within $\sim$20~min as characterized by the materials longitudinal relaxation rate $T_1$ (Fig.~\ref{fig:p and temp}). This result demonstrates that the radiation-hardened bypass actuator can provide stable closed-loop pressure control for repeated TE calibrations, supporting the remote-instrumentation architecture described in Section~\ref{Control}. The polarized target calibration details extracted from these runs will be reported in subsequent publications.

    \subsection{Pressure-Relief System Performance}
    \label{sec:relief_performance}

    The pressure-relief system performed as designed during all quench-test events encountered in commissioning. No relief set points were exceeded, and no pressure-boundary component showed structural failure. The relief network sized by the methodology in Section~\ref{code} therefore satisfied its design basis under transient operating conditions and off-normal magnet-quench events. A detailed analysis of quench response will be reported in a dedicated publication~\cite{quench}.

    \subsection{Demonstrating performance under high power DNP}
    \label{sec:dnp_heatload}

\begin{figure}[h!]
        \centering
        \includegraphics[width=\columnwidth]{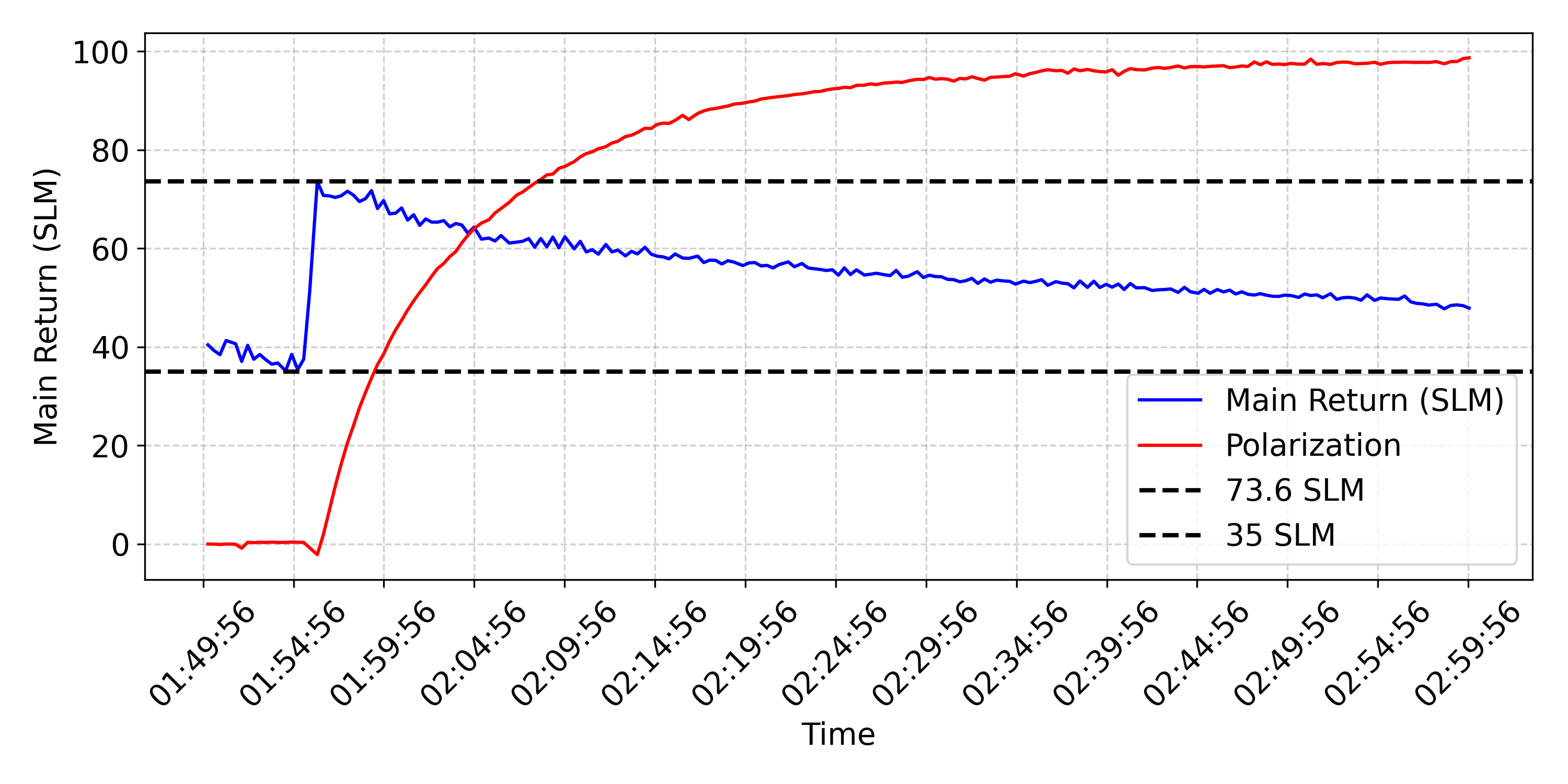}
        \caption{NH\textsubscript{3} polarization above 95\% during high-microwave-power operation, with gradual microwave attenuation applied after initial polarization.}
        \label{fig:polarized}
    \end{figure}

    During DNP enhancement of NH\textsubscript{3}, the refrigerator maintained $\sim$1~K operation while the target was simultaneously polarized and exposed to beam. A proton polarization above 95\% was achieved. Initial polarization runs at maximum microwave power increased helium flow by about 40~SLM (2.5 W). Because these runs were already operating at elevated flow during run-valve configuration tests, attenuation adjustments produced a transient peak flow of 73.6~SLM (Fig.~\ref{fig:polarized}). In this configuration, 80\% proton polarization was reached in the NH3\textsubscript{3} sample in approximately 15~minutes, which is about 5 minutes faster than any previously published ramp up rate for a 1~K 5 T DNP system.

    Variable microwave attenuation can reduce helium consumption while preserving target polarization: high microwave power is applied when rapid polarization is needed, and reduced power is used afterward to conserve liquid helium.

    \subsection{Cooldown Performance}
    \label{sec:cooldown}

\begin{figure}[h!]
        \centering
        \includegraphics[width=\columnwidth]{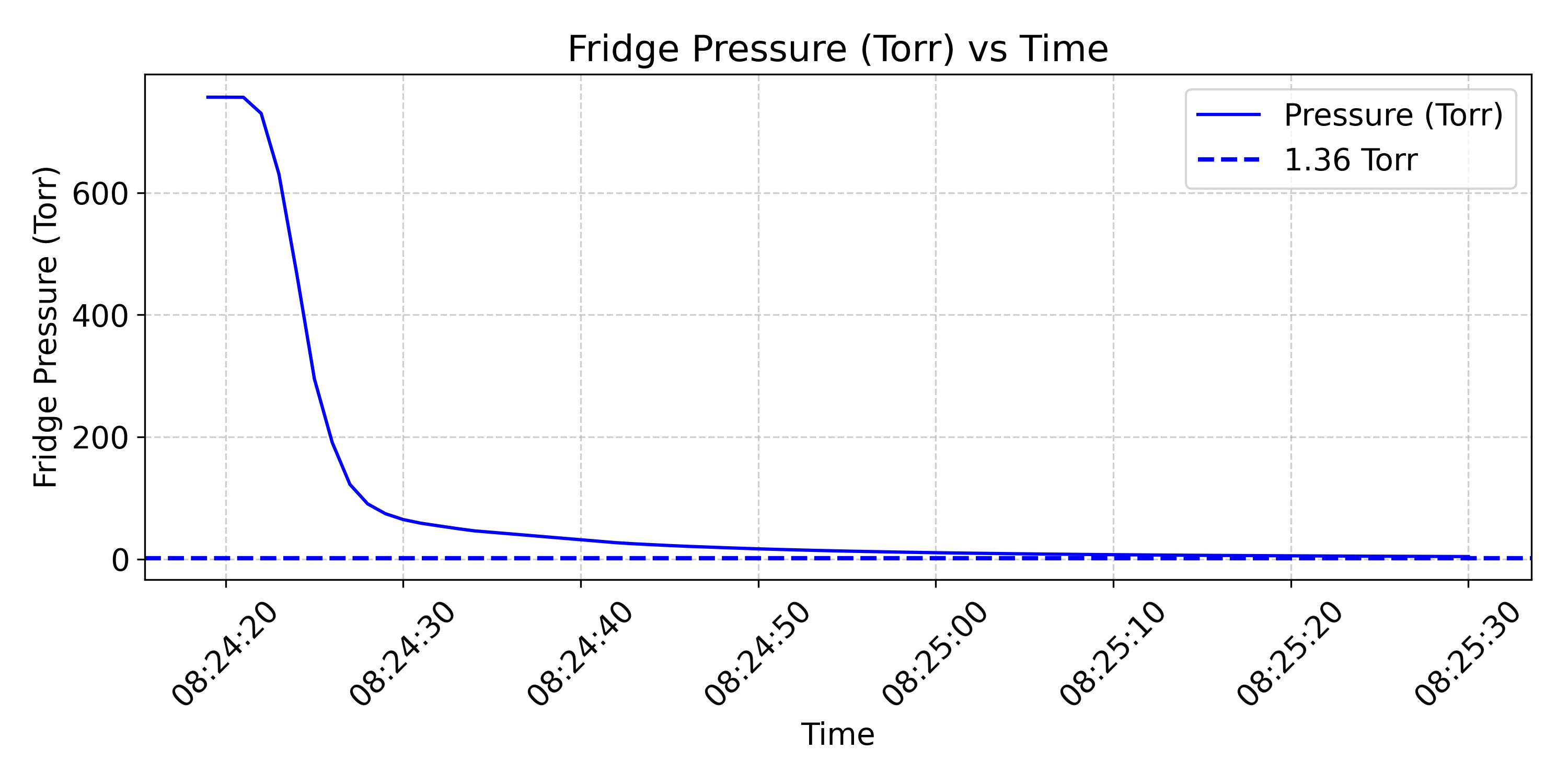}
        \caption{Cooldown profile from ambient temperature to 1.3~K and 1.3~Torr. Total cooldown time was 4~min 14~s.}
        \label{fig:total_cooldown}
    \end{figure}

    Starting from atmospheric pressure, the refrigerator reached 20~Torr in 26~s and 1.3~Torr in 4~min 14~s (Fig.~\ref{fig:total_cooldown}). The rapid cooldown reflects the high effective pumping speed and low parasitic heat load achieved by the radiation-shielded baffle design discussed in Section~\ref{baffle}.

    \subsection{Discussion}
    \label{sec:discussion}

The system met all thermal-performance objectives during commissioning, demonstrating stable operation near 1~K under a maximum operating heat load of approximately 3~W at the highest beam-intensity and microwave-power settings. During routine operation, the heat load was typically kept well below 1~W by increasing microwave attenuation to improve helium efficiency. Under these conditions, stable NH\textsubscript{3} polarizations above 95\% were achieved. TE calibrations were performed using stable PID pressure regulation through the radiation-hardened gate-valve bypass actuator.

The estimated theoretical maximum cooling power, after subtracting the measured baseline heat load, is approximately 8.5 W at operating temperature. The maximum transient heat load from the microwaves demonstrates that the larger DNP load can be fully absorbed while maintaining $\sim$1 K operation. However, the theoretical maximum cooling capacity of this refrigerator was not tested.

The central practical contribution of this work extends beyond FNAL. The compliance methodology is built on national-consensus standards, ASME BPVC Section~VIII for vessels~\cite{asme81} and ASME B31.3 for process piping~\cite{b31}, which are invoked with laboratory-specific overlays by FESHM at FNAL and by corresponding ES\&H frameworks at Jefferson Lab, BNL, and other DOE laboratories with polarized-target programs. A safety basis organized around conservative allowable stresses from low-temperature material data, traceable pressure-boundary fabrication and inspection, and relief sizing keyed to bounding loss-of-vacuum and quench scenarios is therefore portable across these frameworks; only the laboratory-specific review and approval steps need to be remapped. The radiation-tolerant remote-instrumentation approach is similarly portable: passive sensing lines, radiation-hardened actuators with mechanical position feedback, and shielded transducers placed approximately 20~m away can be deployed in other high-radiation hadron-beam target areas, provided the actuator torque and line-impedance characterization are repeated for the new geometry.

Most importantly, the commissioning data show that this compliance pathway does not impose a functional penalty. Full ASME and FESHM compliance and watt-class cryogenic performance were achieved in the same design, indicating that formal pressure-system compliance and high-performance cryogenic operation need not be competing requirements. The compliance pathway has since been adopted as a reference framework for future cryogenic experiments at FNAL and is structurally compatible with the safety processes used at other national laboratories that operate, or plan to operate, polarized-target systems.

\section{Conclusions}
\label{conclusion}

    This paper presented the design, safety basis, and commissioning performance of the SpinQuest \textsuperscript{4}He evaporation refrigerator. The work demonstrates that a high-power, 1~K evaporation refrigerator can be operated in a high-radiation, high-heat-load fixed-target environment while satisfying ASME BPVC, ASME B31.3, and FESHM requirements. The result is not only a functioning refrigerator for SpinQuest, but also a documented design and approval pathway for future cryogenic and polarized target systems.

    The system addresses the three barriers identified at the outset. First, the pressure-system design translates ASME and FESHM requirements into a traceable methodology for a low-pressure cryogenic refrigerator, including conservative allowable stresses, fabrication and inspection records, engineering notes, and relief sizing for bounding loss-of-vacuum and quench scenarios. Second, the remote-instrumentation architecture places sensitive electronics outside the target cave while retaining stable pressure, flow, temperature, and valve-position control through passive sensing lines and radiation-tolerant actuators. Third, the integrated pumping and relief architecture manages environmental, beam-induced, and DNP microwave heat loads while maintaining operation near 1~K and protecting the system during transient events.

    Commissioning confirmed the practical value of this approach. The measured environmental and beam-induced heat loads were each approximately 0.2~W, TE calibrations were stabilized through the radiation-hardened gate-valve bypass, NH\textsubscript{3} proton polarizations above 95\% were achieved during beam operation, and the relief network performed as designed during quench-test events. These results show that formal pressure-system compliance and watt-class cryogenic performance can be achieved in the same design rather than traded against one another.

    The methodology developed here has been adopted as a reference framework for future FNAL cryogenic experiment systems and is being adapted for a planned \textsuperscript{3}He evaporation refrigerator, where sub-1~K operation will introduce additional design constraints. Future work will quantify long-term PID drift under sustained beam conditions, extend NMR calibration studies to NH\textsubscript{3} and ND\textsubscript{3}, and report the detailed magnet-quench response analysis separately.

\section*{Acknowledgments}
    The authors acknowledge Don Crabb as a mentor and dear friend. The authors also thank Erik Voirin for his help navigating the cryogenic safety review. This work was supported by DOE Contract No.~DE-FG02-96ER40950.

\section*{Funding Data}

    \begin{itemize}
        \item This work was supported by the U.S. Department of Energy, Office of Nuclear Physics, Contract No.~DE-FG02-96ER40950.
    \end{itemize}
    
\section*{Nomenclature}
\noindent
\begin{tabbing}
$\Delta P_{\text{U-Tube}}$ \quad \= pressure drop of the U-Tube, psig \\
$P_i$ \> internal pressure, psia \\
$P_e$ \> external pressure, psia \\
$P_r$ \> relief pressure, psig \\
$P_{set}$ \> set pressure for relief, psig \\
$P_d$ \> differential pressure, psia \\
$P_a$ \> external MAWP, psia \\
$S$ \> maximum allowable stress, ksi \\
$W_{RT}$ \> vessel burst energy, ft-lbf \\
\end{tabbing}


    \bibliographystyle{asmejour} 

    \bibliography{spinquest_cryo} 


\end{document}